\documentclass[sigconf,balance=false]{acmart}
\usepackage{popets}

\setcopyright{popets}
\copyrightyear{YYYY}

\acmYear{YYYY}
\acmVolume{YYYY}
\acmNumber{X}
\acmDOI{XXXXXXX.XXXXXXX}
\acmISBN{}
\acmConference{Proceedings on Privacy Enhancing Technologies}
\usepackage{algorithmic}
\usepackage{graphicx}
\usepackage{stfloats}
\usepackage{textcomp}
\usepackage{xcolor}
\usepackage{graphicx}
\usepackage{caption}
\usepackage{subcaption}  
\def\BibTeX{{\rm B\kern-.05em{\sc i\kern-.025em b}\kern-.08em
    T\kern-.1667em\lower.7ex\hbox{E}\kern-.125emX}}

\usepackage{makecell} 
\usepackage[normalem]{ulem}
\usepackage{rotating}
\usepackage{tablefootnote}
\usepackage{tabularray}
\usepackage{hyperref}
\usepackage{microtype}
\usepackage{graphicx}
\usepackage{subcaption} 
\usepackage{booktabs} 

\usepackage{hyperref}

\usepackage{mathtools}
\usepackage{amsthm}

\usepackage[capitalize,noabbrev]{cleveref}

\theoremstyle{plain}

\theoremstyle{definition}

\theoremstyle{remark}

\usepackage[textsize=tiny]{todonotes}

\usepackage{colortbl} 

\usepackage{nccmath} 
\newenvironment{mbmatrix}{\begin{medsize}\begin{bmatrix}}{\end{bmatrix}\end{medsize}}

\usepackage{multirow}
\newcommand{\mysplit}[1]{%
  \begin{tabular}{@{}c@{}}   
    #1
  \end{tabular}
  }
\usepackage{algorithm}      
\usepackage{caption}
\usepackage{subcaption}
\usepackage{graphicx}
\usepackage{caption}
\usepackage{subcaption}
\usepackage{caption}
\usepackage{subcaption}

\begin{document}

\title[]{Volley Revolver: A Novel Matrix-Encoding Method for Privacy-Preserving Neural Networks (Inference)
}


\author{John Chiang}
\orcid{0000-0003-0378-0607}
\email{john.chiang.smith@gmail.com}
\email{https://github.com/petitioner/HE.CNNinfer} %


\renewcommand{\shortauthors}{John Chiang}

\begin{abstract}
In this work, we propose a novel matrix-based encoding scheme that is particularly suitable for enabling privacy-preserving neural networks, and apply it to convolutional neural network (CNN) inference for handwritten image classification over homomorphically encrypted data.
For two matrices $A$ and $B$ to perform homomorphic multiplication, the main idea behind it, in a simple version, is to encrypt matrix $A$ and the transpose of  matrix $B$ into two ciphertexts respectively. With additional operations, the homomorphic matrix multiplication can be calculated over encrypted matrices efficiently. For the convolution operation,  we in advance span each convolution kernel to a matrix space of the same size as the input image so as to generate several ciphertexts, each of which is later used together with the ciphertext encrypting input images for calculating some of the final convolution results. We accumulate all these intermediate results and thus complete the convolution operation. 

In a public cloud environment with 40 vCPUs, our CNN implementation on the MNIST test dataset requires $\sim$ 287 seconds to compute ten likelihoods for 32 simultaneously encrypted images of size 
$28 \times 28$. The data owner only needs to upload a single ciphertext ($\sim 19.8$ MB) that encrypts these 32 images to the public cloud.
\end{abstract}


\maketitle

\section{Introduction}

\subsection{Background}
Machine learning applied in some specific domains such as health and finance should preserve privacy while processing private or confidential data to make accurate predictions. In this study, we focus on privacy-preserving neural network inference, which aims to outsource a well-trained inference model to a cloud service in order to make predictions on private data. For this purpose, the data should be encrypted first and then sent to the cloud service that should not be capable of having access to the raw data. Compared to other cryptology technologies such as Secure Multi-Party Computation, Homomorphic Encryption (HE) provides the most stringent security for this task.

\subsection{Related Work}

The pioneering work by Gilad-Bachrach et al. \cite{gilad2016cryptonets} introduced $\texttt{CryptoNets}$, the first framework to enable high-throughput CNN inference directly on encrypted data. Subsequent research has extensively refined this foundation to address the inherent constraints of Homomorphic Encryption. For instance, Chabanne et al. \cite{chabanne2017privacy} scaled HE-based inference to deeper architectures by leveraging $\texttt{HElib}$ \cite{halevi2020helib} and employing batch normalization to stabilize polynomial approximations of the $\texttt{ReLU}$ activation. To mitigate computational overhead, $\texttt{Faster CryptoNets}$ \cite{chou2018faster} integrated pruning and quantization techniques, while the Low-Latency CryptoNets ($\texttt{LoLa}$) \cite{brutzkus2019low} introduced specialized data encoding schemes to further reduce inference delay. Furthermore, efficiency in secure outsourced matrix multiplication has been significantly enhanced through novel matrix-packing and encoding strategies \cite{jiang2018secure}. Collectively, these advancements aim to overcome the high depth and complexity limitations typically associated with fully homomorphic neural network evaluation \cite{bourse2018fast, chabanne2017privacy}.

Beyond pure Homomorphic Encryption, several privacy-preserving frameworks leverage Multi-Party Computation (MPC) \cite{barni2006privacy, orlandi2007oblivious} or hybrid HE-MPC protocols to accelerate deep learning tasks. The core intuition behind these hybrid approaches is to delegate computationally intensive linear operations, such as dot-product evaluations, to HE, while offloading non-linear activations (e.g., sigmoid or threshold functions) to MPC-based techniques \cite{demmler2015aby, mohassel2017secureml}. Notable implementations include $\texttt{MiniONN}$ \cite{liu2017oblivious}, which employs SIMD batching for oblivious neural network evaluation, and $\texttt{Chameleon}$ \cite{riazi2018chameleon}, which utilizes a trusted third-party framework to optimize offline-phase computation. These methodologies were further refined in $\texttt{GAZELLE}$ \cite{juvekar2018gazelle}, which leverages advanced ciphertext packing and HE automorphisms to enhance throughput. However, despite their localized efficiency gains, these hybrid protocols inherently suffer from substantial communication overhead and prohibitive network latency, driven by the frequent, high-volume interactions required between the participating parties.

\subsection{Contributions}
In this study, our contributions are in three main parts:
\begin{enumerate}
    \item We introduce a novel data-encoding method for matrix multiplications on encrypted matrices, $\texttt{Volley Revolver}$, which  can be used to multiply matrices of arbitrary shape efficiently. 
    \item We propose a feasible evaluation strategy for convolution operation, by devising an efficient homomorphic algorithm to sum some intermediate results of convolution operations.
    \item We develop some simulated operations on the packed ciphertext encrypting an image dataset as if there were multiple virtual ciphertexts inhabiting it, which provides a compelling new perspective of viewing the dataset as a three-dimensional structure. 
\end{enumerate}

\section{Preliminaries}
Let ``$\oplus$'' and ``$\otimes$''  denote the component-wise addition and multiplication respectively between  ciphertexts encrypting matrices  and the ciphertext $\texttt{ct}.P$ the encryption of a matrix $P$. 
Let $I_{[i][j]}^{(m)}$ represent the single pixel of  the $j$-th element in the $i$-th row of the $m$-th image  from the dataset.
\subsection{Fully Homomorphic Encryption} Homomorphic Encryption is one kind of encryption but has its characteristic in that   over an HE system operations on encrypted data generate ciphertexts encrypting the right results of corresponding operations on plaintext without decrypting the data nor requiring access to the  secret key. Since Gentry~\cite{gentry2009fully} presented the first fully homomorphic encryption scheme, tackling the over three decades problem, much progress has been made on an efficient data encoding scheme for the application of machine learning to HE.  
 Cheon et al.~\cite{ckks2017homomorphic} constructed an HE scheme (CKKS) that can deal with this technique problem efficiently, coming up with a new procedure called $\texttt{rescaling}$ for approximate arithmetic in order to manage the magnitude of plaintext. Their open-source library,   $\texttt{HEAAN}$, like other HE libraries also supports the Single Instruction Multiple Data (aka SIMD) manner~\cite{SmartandVercauteren_SIMD} to encrypt multiple values into a single ciphertext. 
 
 Given the security parameter, $\texttt{HEAAN}$ outputs a secret key $\textit{sk}$, a public key $\textit{pk}$, and other public keys used for operations such as rotation.   For simplicity, we will ignore the $\texttt{rescale}$ operation and deem the following operations to deal with the magnitude of plaintext automatedly. $\texttt{HEAAN}$ has the following functions to support the HE scheme:
\begin{enumerate}
  \item \texttt{Enc}$_\textit{pk}$(${m}$): For the public key $\textit{pk}$ and a message vector ${m}$, $\texttt{HEAAN}$ encrypts the message ${m}$ into a ciphertext $\texttt{ct}$.
  
  \item \texttt{Dec}$_\textit{sk}$($\texttt{ct}$): Using the secret key, this algorithm  returns the message vector  encrypted by the ciphertext $\texttt{ct}$. 
  
  \item \texttt{Add}($\texttt{ct}_1$, $\texttt{ct}_2$): This operation returns a new ciphertext that encrypts the message Dec$_\textit{sk}$($\texttt{ct}_1$) $\oplus$ Dec$_\textit{sk}$($\texttt{ct}_2$).
  
  \item \texttt{Mul}($\texttt{ct}_1$, $\texttt{ct}_2$): This procedure returns a new ciphertext that encrypts the message Dec$_\textit{sk}$($\texttt{ct}_1$) $\otimes$ Dec$_\textit{sk}$($\texttt{ct}_2$).
  
  \item \texttt{cMul}($C$, $\texttt{ct}_2$): This procedure returns a new ciphertext that encrypts the message $C$ $\otimes$ Dec$_\textit{sk}$($\texttt{ct}_2$).
  
  \item \texttt{Rot}($\texttt{ct}$, $l$): This procedure generates  a ciphertext encrypting a new plaintext vector obtained by rotating the the original message vector ${m}$ encrypted by $\texttt{ct}$ to the left by $l$ positions. 
\end{enumerate}

\subsection{Database Encoding Method}
For brevity, we assume that the training dataset has $n$ samples with $f$ features and that the number of slots in a single
ciphertext is at least  $n \times f$. A training dataset is usually organized into a matrix $Z$  
 each row of which  represents an example.  Kim et al.~\cite{IDASH2018Andrey} propose an efficient database encoding method to encrypt this matrix into a single ciphertext in a row-by-row manner.   They provide two basic but important shifting operations  by shifting  $1$ and $f$ positions respectively: the $incomplete$ column shifting and the row shifting. The  matrix obtained from  matrix $Z$  by the $incomplete$ column shifting operation is shown as follows:
 
\begin{align*}
 Z &= 
 \begin{bmatrix}
 z_{[1][1]}    &   z_{[1][2]}   &  \ldots  & z_{[1][f]}   \\
 z_{[2][1]}    &   z_{[2][2]}   &  \ldots  & z_{[2][f]}   \\
 \vdots    &   \vdots   &  \ddots  & \vdots   \\
 z_{[n][1]}    &   z_{[n][2]}   &  \ldots  & z_{[n][f]}   \\
 \end{bmatrix}   \\
 &\xmapsto{\textsl{incomplete} \text{ column shifting} }
 \begin{bmatrix}
 z_{[1][2]}    &   z_{[1][3]}   &  \ldots  & z_{[2][1]}   \\
 z_{[2][2]}    &   z_{[2][3]}   &  \ldots  & z_{[3][1]}   \\
 \vdots    &   \vdots   &  \ddots  & \vdots   \\
 z_{[n][2]}    &   z_{[n][3]}   &  \ldots  & z_{[1][1]}   \\
 \end{bmatrix} .  
\end{align*} 

 Han et al.~\cite{han2018efficient} summarize another  two procedures, $\texttt{SumRowVec}$ and $\texttt{SumColVec}$, to compute  the summation of each row and column respectively. The results of two procedures on $Z$ are as follows:
\begin{align*}
\texttt{SumRowVec(${Z}$)} &= \\
&\hspace{-1.2cm}
\begin{bmatrix}
 \sum_{i=1}^n z_{[i][1]}    &   \sum_{i=1}^n z_{[i][2]}   &  \ldots  & \sum_{i=1}^n z_{[i][f]}   \\
 \sum_{i=1}^n z_{[i][1]}    &   \sum_{i=1}^n z_{[i][2]}   &  \ldots  & \sum_{i=1}^n z_{[i][f]}   \\
 \vdots    &   \vdots   &  \ddots  & \vdots   \\
 \sum_{i=1}^n z_{[i][1]}    &   \sum_{i=1}^n z_{[i][2]}   &  \ldots  & \sum_{i=1}^n z_{[i][f]}   \\
\end{bmatrix}, \\[1ex]
\texttt{SumColVec(${Z}$)} &= \\
&\hspace{-1.2cm}
\begin{bmatrix}
 \sum_{j=1}^f z_{[1][j]}    &   \sum_{j=1}^f z_{[1][j]}    &  \ldots  & \sum_{j=1}^f z_{[1][j]}    \\
 \sum_{j=1}^f z_{[2][j]}    &   \sum_{j=1}^f z_{[2][j]}   &  \ldots  & \sum_{j=1}^f z_{[2][j]}   \\
 \vdots    &   \vdots   &  \ddots  & \vdots   \\
 \sum_{j=1}^f z_{[n][j]}    &   \sum_{j=1}^f z_{[n][j]}   &  \ldots  & \sum_{j=1}^f z_{[n][j]}   \\
\end{bmatrix}.
\end{align*}

We propose a new useful procedure called $\texttt{SumForConv}$ to facilitate  convolution operation for every image, as shown in Algorithm \ref{alg:SumForConv}. The computational overhead of $\texttt{SumForConv}$ for a kernel of size $k \times k$ is characterized by $O(2k)$ homomorphic additions, $1$ constant multiplication, and $O(2k)$ ciphertext rotations. This complexity profile ensures that the spatial summation required for the convolution remains efficient, as the rotation count—the primary latency bottleneck—scales linearly with the kernel's dimensions rather than quadratically.  Since rotation operation is comparably expensive than the other two operations, the complexity can be seen as asymptotically $O(k)$ rotations. Below we illustrate the result of $\texttt{SumForConv}$ on $Z$ taking the example that  $n$ and $f$ are both $4$ and  the kernel size is $3 \times 3$:
\begin{align*}
 Z &= 
 \begin{bmatrix}
 z_{[1][1]}    &   z_{[1][2]}   &  z_{[1][3]}  & z_{[1][4]}   \\
 z_{[2][1]}    &   z_{[2][2]}   &  z_{[2][3]}  & z_{[2][4]}   \\
 z_{[3][1]}    &   z_{[3][2]}   &  z_{[3][3]}  & z_{[3][4]}   \\
 z_{[4][1]}    &   z_{[4][2]}   &  z_{[4][3]}  & z_{[4][4]}   \\
 \end{bmatrix} \\
 &\xmapsto{ \texttt{SumForConv}(\cdot,3,3) }
 \begin{bmatrix}
 s_{[1][1]}    &   s_{[1][2]}   &  0       & 0   \\
 s_{[2][1]}    &   s_{[2][2]}   &  0       & 0   \\
 0         &   0        &  0       & 0   \\
 0         &   0        &  0       & 0   \\
 \end{bmatrix},
\end{align*} 
where $s_{[i][j]} = \sum_{p=i}^{i+2} \sum_{q=j}^{j+2} z_{[p][q]} $ for $1 \le i, j \le 2$. In  the convolutional layer,  $\texttt{SumForConv}$ can help to compute some partial results of  convolution operation for an image simultaneously. 
 
\begin{algorithm}[htbp]
    \caption{SumForConv: sum some part results of convolution operation after one element-wise multiplication}
    \label{alg:SumForConv}
    \begin{algorithmic}[1]
        \STATE {\bfseries Require:} a ciphertext $\texttt{ct}.I$ encrypting a (convolved) image $I$ of size $h \times w$, the size $k \times k$ of some kernel $K$ with its bias $k_0$, and a stride of $(1, 1)$;
        \STATE {\bfseries Ensure:} a ciphertext $\texttt{ct}.I_s$ encrypting a resulting image $I_s$ of the same size as $I$

        \STATE {\bfseries Set} $I_s \gets \boldsymbol 0$
        \# $I_s \in \mathbb R^{(h-k+1) \times (w-k+1)}$

        \FOR{$i := 1$ to $(h-k+1)$}
            \FOR{$j := 1$ to $(w-k+1)$}
                \STATE $I_s[i][j] \gets k_0$
            \ENDFOR
        \ENDFOR

        \STATE $\texttt{ct}.I_s \gets \texttt{Enc}_\textit{pk}(I_s)$

        \STATE \# Accumulate columns (could be computed in parallel)
        \FOR{$pos := 0$ to $k-1$}
            \STATE $\texttt{ct}.T \gets \texttt{Rot}(\texttt{ct}.I, pos)$  
            \STATE $\texttt{ct}.I_s \gets \texttt{Add}(\texttt{ct}.I_s, \texttt{ct}.T)$
        \ENDFOR
        
        \STATE \# Accumulate rows (could be computed in parallel)
        \FOR{$pos := 1$ to $k-1$}
            \STATE $\texttt{ct}.T \gets \texttt{Rot}(\texttt{ct}.I, pos \times w)$
            \STATE $\texttt{ct}.I_s \gets \texttt{Add}(\texttt{ct}.I_s, \texttt{ct}.T)$
        \ENDFOR
        
        \STATE \# Build a new designed matrix to filter out the garbage values
        \STATE {\bfseries Set} $M \gets \boldsymbol 0$
        \# $M \in \mathbb R^{h \times w }$

        \FOR{$hth := 0$ to $(h-1)$}
            \FOR{$wth := 0$ to $(w-1)$}
                \IF{$wth \bmod k = 0 $ {\bfseries and} $wth + k \leq width$ {\bfseries and} $hth \bmod k = 0$ {\bfseries and} $hth + k \leq height$}
                    \STATE $M[hth][wth] \gets 1$
                \ENDIF 
            \ENDFOR
        \ENDFOR

        \STATE $\texttt{ct}.I_s \gets \texttt{cMul}(M, \texttt{ct}.I_s)$
 
        \STATE {\bfseries Return} $\texttt{ct}.I_s$
    \end{algorithmic}
\end{algorithm}

\subsection{Convolutional Neural Networks }
Convolutional Neural Networks are neural networks particularly tailored for image recognition, equipped with two distinct kinds of layers:  Convolutional layer (CONV) and Pooling layer (POOL) in addition to another two basic kinds of layers: Fully Connected layer and Activation layer (ACT). A CNN for image classification  has a common  architecture:
$[[CONV \to ACT]^p \to POOL]^q \to [CONV \to ACT] \to [FC \to ACT]^r \to FC$,
where $p$, $q$ and $r$  are integers usually greater than $1$. In our implementation, we use the same CNN architecture as \cite{jiang2018secure}: $[CONV \to ACT] \to [FC \to ACT] \to FC$. 

Convolutional layer is the fundamental basis of a CNN, which has kernels of size $k \times k$, a stride of $(s, s)$, and a channel (mapcount) of $c$. Each kernel has $k \times k \times c$ trainable parameters, along with the kernel bias $k_0$, all of which are updated during the training process. Given a greyscale image $I \in \mathbb R^{h \times w}$ and a kernel $K \in \mathbb R^{k \times k}$, the result of  convolving this input image $I$ with stride of $(1, 1)$ is the output image $I^{\prime} \in \mathbb R^{h^{\prime} \times w^{\prime}}$ with $I_{[i^{\prime}][j^{\prime}]}^{\prime} = k_0 + \sum_{i = 1}^k { \sum_{j = 1}^k { K_{[i][j]} \times I_{[i^{\prime}+i][j^{\prime}+j]} } } $ for  $ 0 <  i^{\prime} \le h-k+1 $ and $ 0 < j^{\prime} \le w-k+1 $. It can be extended to a color image or a  convolved image with many channels, refer to [9] for a detail. If there are multiple kernels, the convolutional layer stacks all the convolved results of each kernel. For a single input sample, FC layer only accepts a unidimensional vector, which is why the output of previous layers ($[[CONV \to ACT]^p \to POOL]$ or $[CONV \to ACT]$) should be flattened before being fed into FC layer.

\section{Technical Details}
We introduce a novel matrix-encoding method called  $\texttt{Volley Revolver}$, which is particularly suitable for secure matrix multiplication. 
 The core idea is to place each semantically-complete  information (such as an example in a dataset) into the corresponding  row of a matrix and encrypt this matrix into a single ciphertext. 
  When applying it to private neural networks, $\texttt{Volley Revolver}$ puts the whole weights of every neural node into the corresponding row of a matrix, organizes all the nodes from the same layer into this matrix, and encrypts this matrix into a single ciphertext.

\subsection{Encoding Method for Matrix Multiplication}
\label{section:EncodingMethodforMatrixMultiplication}
Suppose we are given an $m \times n$ matrix $A$ and an $n \times p$ matrix $B$. Let $C$ be the $m \times p$ matrix defined as their product, $C = A \cdot B$, with elements $C_{[i][j]} = \sum_{k=1}^n a_{[i][k]} \times b_{[k][j]} $: 
\begin{align*}
  A &= 
 \begin{bmatrix}
 a_{[1][1]}    &   a_{[1][2]}   &  \ldots  & a_{[1][n]}   \\
 a_{[2][1]}    &   a_{[2][2]}   &  \ldots  & a_{[2][n]}   \\
 \vdots    &   \vdots   &  \ddots  & \vdots   \\
 a_{[m][1]}    &   a_{[n][2]}   &  \ldots  & a_{[m][n]}   \\
 \end{bmatrix} 
 ,  \\
 B &= 
 \begin{bmatrix}
 b_{[1][1]}    &   b_{[1][2]}   &  \ldots  & b_{[1][p]}   \\
 b_{[2][1]}    &   b_{[2][2]}   &  \ldots  & b_{[2][p]}   \\
 \vdots    &   \vdots   &  \ddots  & \vdots   \\
 b_{[n][1]}    &   b_{[n][2]}   &  \ldots  & b_{[n][p]}   \\
 \end{bmatrix}.
\end{align*}   
For simplicity, we assume that each of the three matrices ${A}$, ${B}$ and ${C}$ could be encrypted into a single ciphertext. We also make the assumption that  $m$ is greater than $p$, $m > p$. We will not  illustrate the other cases where $m \le p$, which is similar to this one. When it comes to the homomorphic matrix multiplication,  $\texttt{Volley Revolver}$ encodes matrix $A$ directly but  encodes  the padding form of the transpose of  matrix $B$, by using two row-ordering encoding maps. For matrix ${A}$, we adopt the same encoding method that \cite{jiang2018secure} did by the encoding map $\tau_a : {A} \mapsto {\bar{A}} = (a_{[1 + (k/n)][1 + (k\%n)]})_{0 \le k < m \times n}.$ 
For  matrix ${B}$, we design a very different encoding method from \cite{jiang2018secure} for $\texttt{Volley Revolver}$ : We transpose the matrix $B$ first, and then tile the resulting matrix vertically to obtain a matrix of size $m \times n$. Therefore  $\texttt{Volley Revolver}$ adopts the encoding map $\tau_b : {B} \mapsto {\bar{B}} = (b_{[1 + (k\%n)][1 + ((k/n)\%p)]})_{0 \le k < m \times n}$, obtaining  the matrix from  mapping $\tau_b$ on $B$: 
\begin{align*}
 B &= 
 \begin{bmatrix}
 b_{[1][1]}    &   b_{[1][2]}   &  \ldots  & b_{[1][p]}   \\
 b_{[2][1]}    &   b_{[2][2]}   &  \ldots  & b_{[2][p]}   \\
 \vdots    &   \vdots   &  \ddots  & \vdots   \\
 b_{[n][1]}    &   b_{[n][2]}   &  \ldots  & b_{[n][p]}   \\
 \end{bmatrix}  
 \xmapsto{\tau_b }  \\
 &\begin{bmatrix}
 b_{[1][1]}    &   b_{[2][1]}   &  \ldots  & b_{[n][1]}   \\
 b_{[1][2]}    &   b_{[2][2]}   &  \ldots  & b_{[n][2]}   \\
 \vdots    &   \vdots   &  \ddots  & \vdots   \\
 b_{[1][p]}    &   b_{[2][p]}   &  \ldots  & b_{[n][p]}   \\
 b_{[1][1]}    &   b_{[2][1]}   &  \ldots  & b_{[n][1]}   \\
 \vdots    &   \vdots   &  \ddots  & \vdots   \\
 b_{[1][1 + (m-1)\%p]}    &   b_{[2][1 + (m-1)\%p]}   &  \ldots  & b_{[n][1 + (m-1)\%p]}   \\
 \end{bmatrix}.
\end{align*}

\subsection{Homomorphic Matrix Multiplication}
We report an efficient evaluation algorithm for homomorphic matrix multiplication. This algorithm uses a ciphertext $\texttt{ct}.R$, which encrypts zeros or a given value (such as the weight bias of a fully-connected layer), as an accumulator, and an operation $\texttt{RowShifter}$ to perform a specific type of row shifting on the encrypted matrix ${\bar{B}}$. $\texttt{RowShifter}$ removes the first row of ${\bar{B}}$ and appends a corresponding existing row from ${\bar{B}}$.
\begin{align*}
 {\bar{B}} &= 
 \begin{bmatrix}
 b_{[1][1]}    &   b_{[2][1]}   &  \ldots  & b_{[n][1]}   \\
 b_{[1][2]}    &   b_{[2][2]}   &  \ldots  & b_{[n][2]}   \\
 \vdots    &   \vdots   &  \ddots  & \vdots   \\
 b_{[1][p]}    &   b_{[2][p]}   &  \ldots  & b_{[n][p]}   \\
 b_{[1][1]}    &   b_{[2][1]}   &  \ldots  & b_{[n][1]}   \\
 \vdots    &   \vdots   &  \ddots  & \vdots   \\
 b_{[1][r]}    &   b_{[2][r]}   &  \ldots  & b_{[n][r]}   \\
 \end{bmatrix}
 \xmapsto{\texttt{RowShifter}( {\bar B} )}  \\
 &\begin{bmatrix}
 b_{[1][2]}    &   b_{[2][2]}   &  \ldots  & b_{[n][2]}   \\
 \vdots    &   \vdots   &  \ddots  & \vdots   \\
 b_{[1][p]}    &   b_{[2][p]}   &  \ldots  & b_{[n][p]}   \\
 b_{[1][1]}    &   b_{[2][1]}   &  \ldots  & b_{[n][1]}   \\
 \vdots    &   \vdots   &  \ddots  & \vdots   \\
 b_{[1][r]}    &   b_{[2][r]}   &  \ldots  & b_{[n][r]}   \\
 b_{[1][(r+1)\%p]}    &   b_{[2][(r+1)\%p]}   &  \ldots  & b_{[n][(r+1)\%p]}   \\
 \end{bmatrix} .
\end{align*} 
Algorithm \ref{alg:RowShifter} describes how the procedure $\texttt{RowShifter}$  generates a new ciphertext  from $\texttt{ct}.\bar{B}$.
\begin{algorithm}[htbp]
    \caption{RowShifter: To shift row like a revolver}
    \label{alg:RowShifter}
    \begin{algorithmic}[1]
        \STATE {\bfseries Require:} a ciphertext $\texttt{ct}.M$ encrypting a matrix $M$ of size $m \times n$, the number $p$, and the number $idx$ that is determined in Algorithm \ref{alg:Homomorphic_matrix_multiplication}
        \STATE {\bfseries Ensure:} a ciphertext $\texttt{ct}.R$ encrypting the resulting matrix $R$ of the same size as $M$

        \STATE {\bfseries Set} $R \gets \boldsymbol 0$
        \# $R \in \mathbb R^{m \times n}$
        \STATE $\texttt{ct}.R \gets \texttt{Enc}_\textit{pk}(R)$
        
        \STATE \# Step 1: Rotate the ciphertext first and then filter out the last row
        \STATE $\texttt{ct}.T \gets \texttt{Rot}(\texttt{ct}.M, n)$
        
        \STATE \# Build a specially designed matrix to filter out the last row
        \STATE {\bfseries Set} $F_1 \gets \boldsymbol 1$
        \# $F_1 \in \mathbb R^{m \times n}$
        \FOR{$j := 1$ to $n$}
            \STATE $F_1[m][j] \gets 0$
        \ENDFOR
        \STATE $\texttt{ct}.T_1 \gets \texttt{cMul}(F_1, \texttt{ct}.T)$
          
        \STATE \# Step 2: Rotate the ciphertext first and then filter out the last row
        \STATE $\texttt{ct}.P \gets \texttt{Rot}(\texttt{ct}.M, n \times ((m \bmod p + idx + 1) \bmod p - idx))$
        
        \STATE \# Build a specially designed matrix to filter out the last row
        \STATE {\bfseries Set} $F_2 \gets \boldsymbol 0$
        \# $F_2 \in \mathbb R^{m \times n}$
        \FOR{$j := 1$ to $n$}
            \STATE $F_2[m][j] \gets 1$
        \ENDFOR
        \STATE $\texttt{ct}.T_2 \gets \texttt{cMul}(F_2, \texttt{ct}.P)$
          
        \STATE \# Concatenate            
        \STATE $\texttt{ct}.R \gets \texttt{Add}(\texttt{ct}.T_1, \texttt{ct}.T_2)$
 
        \STATE {\bfseries Return} $\texttt{ct}.R$
    \end{algorithmic}
\end{algorithm}
 
Given two ciphertexts, $\texttt{ct}.A$ and $\texttt{ct}.{\bar{B}}$, the homomorphic matrix multiplication algorithm proceeds in $p$ iterations. During the $k$-th iteration ($0 \le k < p$), it performs the following four steps:

\indent $\texttt{ Step 1:}$ 
In this step, the $\texttt{RowShifter}$ operator is applied to $\texttt{ct}.\bar{B}$ to generate a rotated ciphertext $\texttt{ct}.\bar{B}_1$. Subsequently, a homomorphic multiplication is performed between $\texttt{ct}.A$ and $\texttt{ct}.\bar{B}_1$, yielding the product ciphertext $\texttt{ct}.A\bar{B}_1$. Note that in the trivial case where $k = 0$, $\texttt{RowShifter}$ simply returns an identity copy of $\texttt{ct}.\bar{B}$.

\indent $\texttt{ Step 2:}$ 
In this stage, the public cloud invokes the $\texttt{SumColVec}$ operator on $\texttt{ct}.A\bar{B}_1$ to compute the row-wise summation of $A\bar{B}_1$. The resulting sums are then broadcast across their respective rows to populate the entire matrix, yielding the transformed ciphertext $\texttt{ct}.D$.

\indent $\texttt{ Step 3:}$ This step designs a special matrix ${F}$ for filtering out the redundancy element in $D$ by one  constant multiplication \texttt{cMul}(${F}$, $\texttt{ct}.{D}$), producing the ciphertext $\texttt{ct}.{D}_1$.  

\indent $\texttt{ Step 4:}$ The ciphertext $\texttt{ct}.R$ is then used to accumulate the intermediate  ciphertext $\texttt{ct}.{D}_1$.  

The algorithm repeats Steps 1 to 4 for $p$ iterations and finally aggregates all intermediate ciphertexts, returning the ciphertext $\texttt{ct}.C$. Algorithm \ref{alg:Homomorphic_matrix_multiplication} shows how to perform our homomorphic matrix multiplication.   
Table~\ref{tab2} summarizes the complexity of each step of Algorithm~\ref{alg:Homomorphic_matrix_multiplication}. 
\begin{table}[bht]
\centering
\caption{ Complexity and required depth of Algorithm~\ref{alg:Homomorphic_matrix_multiplication} }
\label{tab2}
\begin{tabular}{|c||c|c|c|c|c|}
\hline
$\texttt{Step}$     & $\texttt{Add}$  & $\texttt{cMult}$  & $\texttt{Rot}$  & $\texttt{Mult}$     & $\texttt{Depth}$  \\
\hline\hline
$\texttt{1}$        &   1       &  2       &   2       &   1      &   \begin{tabular}{@{}c@{}} 1 CMult \\ + \\ 1 Mult \end{tabular}        \\
\hline
$\texttt{2}$        &    2 $log_2 p$    &   1       &  2$ \log_2 p$      & 0       &  1 CMult  \\
\hline
$\texttt{3}$        &  0   &  1       &  0       &  0        &   1 CMult          \\
\hline
$\texttt{4}$        &  1       &  0       &   0       &  0         &   0       \\
\hline\hline
$\texttt{Total}$    &   $O(p\log p)$    &   $O(p)$    &   $O(p\log p)$       &    $O(p)$            &  \begin{tabular}{@{}c@{}} 1 Mult \\ + \\ 3 CMult \end{tabular}      \\
\hline
\end{tabular}
\end{table}

\begin{figure*}[htp]
\centering
\includegraphics[scale=0.6]{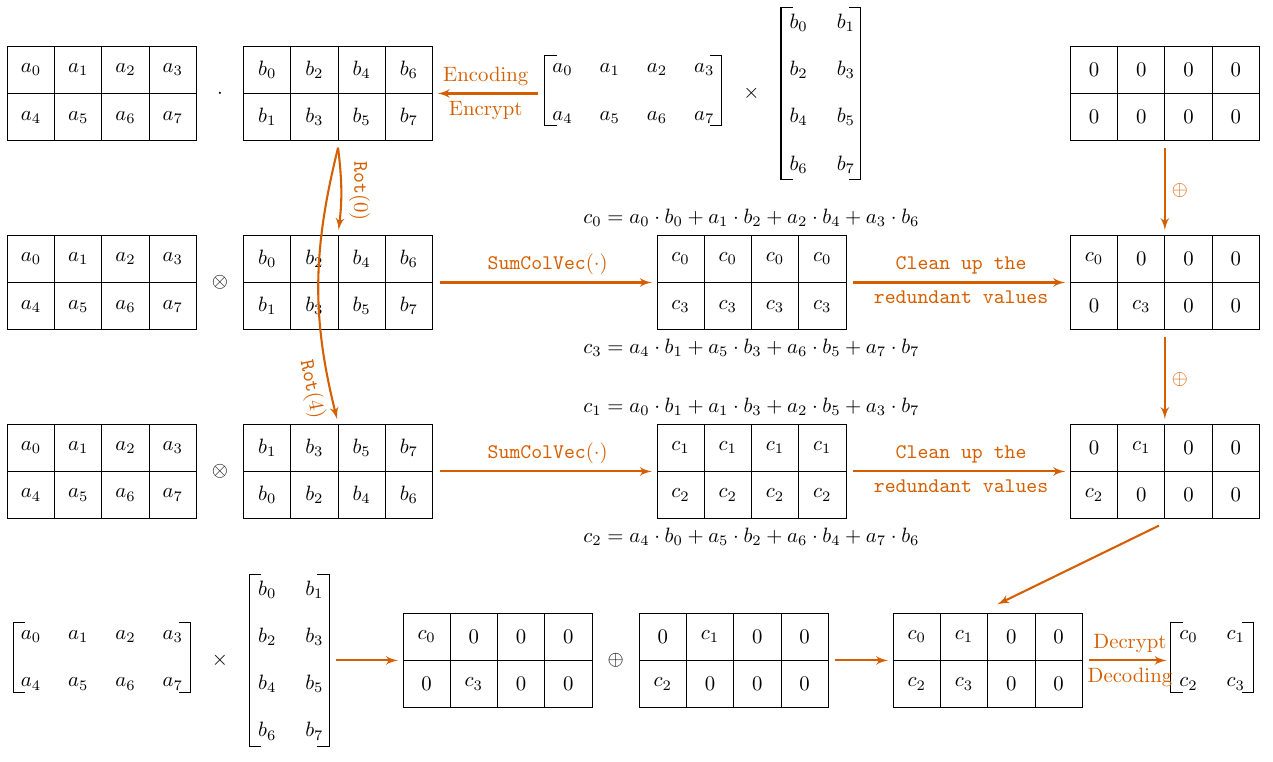}
\caption{
 Our matrix multiplication algorithm with $m = 2$, $n = 4$ and $p = 2$}
\label{ Matrix Multiplication }
\end{figure*}

Figure \ref{ Matrix Multiplication }  describes a simple case for  Algorithm \ref{alg:Homomorphic_matrix_multiplication} where $m = 2$, $n = 4$ and $p = 2$.  

\begin{algorithm}[htbp]
    \caption{Homomorphic matrix multiplication}
    \label{alg:Homomorphic_matrix_multiplication}
    \begin{algorithmic}[1]
        \STATE {\bfseries Require:} $\texttt{ct}.A$ and $\texttt{ct}.{\bar{B}}$ for $A \in \mathbb{R}^{m \times n}$, $B \in \mathbb{R}^{n \times p}$ and $B \xmapsto{\text{Volley Revolver Encoding}} {\bar{B}} \in \mathbb{R}^{m \times n}$;
        \STATE {\bfseries Ensure:} The encrypted resulting matrix $\texttt{ct}.C$ for $C \in \mathbb{R}^{m \times p}$ of the matrix product $A \cdot B$

        \STATE {\bfseries Set} $C \gets \boldsymbol 0$
        \# $C$: To accumulate intermediate matrices 
        \STATE $\texttt{ct}.C \gets \texttt{Enc}_\textit{pk}(C)$
        
        \STATE \# The outer loop (could be computed in parallel)
        \FOR{$idx := 0$ to $p-1$}
            \STATE $\texttt{ct}.T \gets \texttt{RowShifter}(\texttt{ct}.{\bar{B}}, p, idx)$
            \STATE $\texttt{ct}.T \gets \texttt{Mul}(\texttt{ct}.A, \texttt{ct}.T)$
            \STATE $\texttt{ct}.T \gets \texttt{SumColVec}(\texttt{ct}.T)$  
            
            \STATE \# Build a specifically-designed matrix to clean up the redundant values            
            \STATE {\bfseries Set} $F \gets \boldsymbol 0$
            \# $F \in \mathbb R^{m \times n}$
            \FOR{$i := 1$ to $m$}
                \STATE $F[i][(i+idx) \bmod n] \gets 1$
            \ENDFOR
            \STATE $\texttt{ct}.T \gets \texttt{cMul}(F, \texttt{ct}.T)$           
            
            \STATE \# To accumulate the intermediate results
            \STATE $\texttt{ct}.C \gets \texttt{Add}(\texttt{ct}.C, \texttt{ct}.T)$    
        \ENDFOR
          
        \STATE {\bfseries Return} $\texttt{ct}.C$
    \end{algorithmic}
\end{algorithm}

The computation process of this method, particularly in the simple case where $m = p$, is intuitively similar to a special type of revolver that can fire multiple bullets simultaneously (with the first matrix ${A}$ held fixed while the second matrix ${B}$ rotates). This analogy motivates the name of our encoding method, “$\texttt{Volley Revolver}$.” In real-world cases where $m \bmod p = 0$, the $\texttt{RowShifter}$ operation reduces to a single rotation, $\texttt{RowShifter} = \texttt{Rot}(\texttt{ct}, n)$, which is  more efficient and should therefore be adopted whenever possible. For neural networks, this can be achieved by setting the number of nodes in each fully-connected layer to a power of two.

\subsection{Homomorphic  Convolution Operation}
In this subsection, we first introduce a novel but impractical algorithm to calculate the convolution operation for a single grayscale image of size $h \times w$ based on the assumption that this single image can $happen$ to be encrypted into a single ciphertext without vacant slots left, meaning the number $N$ of slots in a packed ciphertext chance to be  $N = h \times w$. We then illustrate how to use this method to compute the convolution operation of several images of $any ~ size$ at the same time  for a convolutional layer after these images have been encrypted into a ciphertext and been viewed as several virtual ciphertexts inhabiting this real ciphertext. For simplicity, we assume that the image is grayscale and that the image dataset can be encrypted into a single ciphertext. 

\textbf{An impractical algorithm }
Given a grayscale image $I$ of size $h \times w$  and a kernel $K$ of size $k \times k$ with its bias $k_0$ such that $h$ and $w$ are both greater than $k$, based on the assumption that this image can $happen$ to be encrypted into a ciphertext $\texttt{ct}.I$ with no more or less vacant slots, we present an efficient algorithm to compute the convolution operation. We set the stride size to the usual default value $(1, 1)$ and adopt  no padding technique  in this algorithm.

Before the algorithm starts, the kernel $K$ should be called by an operation that we term $\texttt{Kernelspanner}$ to  in advance generate $k^2$  ciphertexts for most cases where $h \ge 2\cdot k -1$ and $w \ge 2\cdot k -1$, each of which encrypts a  matrix $P_i$ for $ 1 \le i \le k^2  $ , using a map to span the $k \times k$ kernel to a $h \times w$ matrix space. 
For a simple example that $h = 4$, $w = 4$ and $k = 2$, $\texttt{Kernelspanner}$ generates four ciphertexts, and the kernel bias $k_0$ is used to generate an additional ciphertext:
\begin{align*}
&
\begin{bmatrix}
 k_{1} & k_{2} \\
 k_{3} & k_{4}
\end{bmatrix}
\xmapsto[\mathbb{R}^{k \times k} \mapsto k^2 \cdot \mathbb{R}^{h \times w}]{\texttt{Kernelspanner}} \\
&\quad\quad\quad
\begin{aligned}
&Enc\begin{bmatrix}
 k_{1} & k_{2} & k_{1} & k_{2} \\
 k_{3} & k_{4} & k_{3} & k_{4} \\
 k_{1} & k_{2} & k_{1} & k_{2} \\
 k_{3} & k_{4} & k_{3} & k_{4}
\end{bmatrix},
&&
Enc\begin{bmatrix}
 0 & k_{1} & k_{2} & 0 \\
 0 & k_{3} & k_{4} & 0 \\
 0 & k_{1} & k_{2} & 0 \\
 0 & k_{3} & k_{4} & 0
\end{bmatrix},
\\[1em]
&Enc\begin{bmatrix}
 0 & 0 & 0 & 0 \\
 0 & k_{1} & k_{2} & 0 \\
 0 & k_{3} & k_{4} & 0 \\
 0 & 0 & 0 & 0
\end{bmatrix},
&&
Enc
\begin{bmatrix}
 0 & 0 & 0 & 0 \\
 k_{1} & k_{2} & k_{1} & k_{2} \\
 k_{3} & k_{4} & k_{3} & k_{4} \\
 0 & 0 & 0 & 0
\end{bmatrix}.
\end{aligned}
\\[1em]
&
\begin{bmatrix}
 k_{0}
\end{bmatrix}
\mapsto
Enc
\begin{bmatrix}
 k_{0} & k_{0} & k_{0} & 0 \\
 k_{0} & k_{0} & k_{0} & 0 \\
 k_{0} & k_{0} & k_{0} & 0 \\
 0 & 0 & 0 & 0
\end{bmatrix}.
&
\end{align*}

Our preliminary homomorphic algorithm for convolution operation also needs a ciphertext $\texttt{ct}.R$ to accumulate the intermediate ciphertexts, which should be initially encrypted by the kernel bias $k_0$. This algorithm requires  $k \times k$ iterations and the $i$-th iteration consists of the following four steps for $1 \le i \le k^2$: 

\indent $\texttt{ Step 1:}$ For  ciphertexts $\texttt{ct}.I$ and $\texttt{ct}.P_i$, this step computes their multiplication and returns the  ciphertext  $\texttt{ct}.{IP_i}  = \texttt{Mul}(\texttt{ct}.I, \texttt{ct}.P_i)$. 

\indent $\texttt{ Step 2:}$ To aggregate the values of some  blocks of size $k \times k$, this step applies the procedure $\texttt{SumForConv}$   on the ciphertext $\texttt{ct}.{IP_i}$, producing the ciphertext $\texttt{ct}.{D}$. 

\indent $\texttt{ Step 3:}$ The public cloud generates a specially-designed matrix in order to filter out the garbage data in $\texttt{ct}.{D}$ by one constant multiplication,  resulting in the ciphertext $\texttt{ct}.{\bar D}$. 

\indent $\texttt{ Step 4:}$ In this step, the homomorphic convolution-operation algorithm   updates the  accumulator ciphertext $\texttt{ct}.{R}$ by homomorphically adding $\texttt{ct}.{\bar D}$ to it, namely $\texttt{ct}.{R}  = \texttt{Add}(\texttt{ct}.R, \texttt{ct}.{\bar D})$.  

Note that Steps 1--3 in this algorithm can be computed in parallel with $k \times k$ threads. We describe how to compute homomorphic convolution operation in Algorithm~\ref{alg:Homomorphic_convolution_operation} in detail.  
Table~\ref{tab3} summarizes the total complexity of Algorithm~\ref{alg:Homomorphic_convolution_operation}. 
\begin{table}[bht]
\centering
\caption{ Complexity and required depth of Algorithm~\ref{alg:Homomorphic_convolution_operation} }
\label{tab3}
\begin{tabular}{|c||c|c|c|c|c|}
\hline
$\texttt{Step}$     & $\texttt{Add}$  & $\texttt{cMult}$  & $\texttt{Rot}$  & $\texttt{Mult}$   & $\texttt{Depth}$   \\
\hline\hline
$\texttt{1}$        &   0      &   0       &   0       &   1    &   1 Mult        \\
\hline
$\texttt{2}$        &   $2k$      &   1       &   $2k$       &   0      &   1 CMult      \\
\hline
$\texttt{3}$        &  0   &   1       &   0       &   0       &   1 CMult        \\
\hline
$\texttt{4}$        &  1       &   0       &   0       &   0       &   0        \\
\hline\hline
$\texttt{Total}$    &  $O(k^3)$    &   $O(k^2)$       &  $O(k^3)$      &    $O(k^2)$          &   \makecell[c]{1 Mult \\ + \\ 2 CMult}          \\
\hline
\end{tabular}
\end{table}

\begin{figure}[htp]
\centering
\includegraphics[scale=.6]{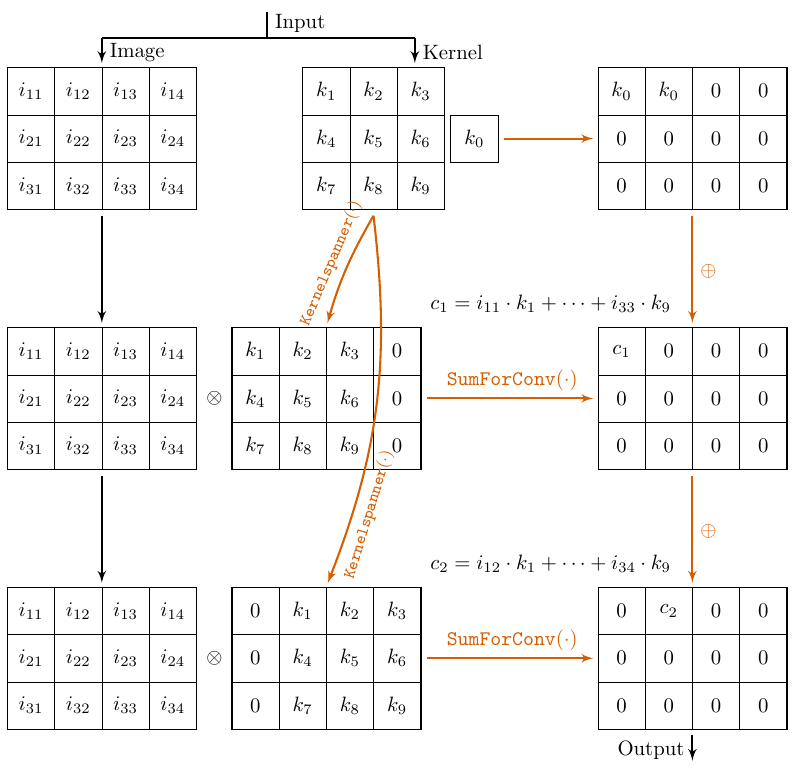}
\caption{
 Our convolution operation algorithm with $h = 3$, $w = 4$ and $k = 3$}
\label{ convolution calculation }
\end{figure}

Figure  \ref{ convolution calculation }  describes a simple case for the algorithm  where $h = 3$, $w = 4$ and $k = 3$.

\begin{algorithm}[htbp]
    \caption{Homomorphic convolution operation}
    \label{alg:Homomorphic_convolution_operation}
    \begin{algorithmic}[1]
        \STATE {\bfseries Require:} An encrypted Image $\texttt{ct}.I$ for $I \in \mathbb{R}^{h \times w}$ and a kernel $K$ of size $k \times k$ with its bias $k_0$
        \STATE {\bfseries Ensure:} The encrypted resulting image $\texttt{ct}.I_s$ where $I_s$ has the same size as $I$

        \STATE \# The Third Party performs $\texttt{Kernelspanner}$ and prepares the ciphertext encrypting kernel bias 
        \STATE $\texttt{ct}.S_{[i]} \gets \texttt{Kernelspanner}(K, h,w)$
        \# $1 \le i \le k^2$ 
        
        \STATE {\bfseries Set} $I_s \gets \boldsymbol 0$
        \# $I_s \in \mathbb R^{h \times w}$
        \FOR{$i := 1$ to $h - k + 1$}
            \FOR{$j := 1$ to $w - k + 1$}
                \STATE $I_s[i][j] \gets k_0$
            \ENDFOR
        \ENDFOR        
        \STATE $\texttt{ct}.I_s \gets \texttt{Enc}_\textit{pk}(I_s)$ 
        
        \STATE \# So begins the Cloud its work
        \FOR{$i := 0$ to $k-1$}
            \FOR{$j := 0$ to $k-1$}
                \STATE $\texttt{ct}.T \gets \texttt{Mul}(\texttt{ct}.I, \texttt{ct}.S_{[i \times k + j + 1]})$
                \STATE $\texttt{ct}.T \gets \texttt{SumForConv}(\texttt{ct}.T)$ 
                
                \STATE \# Design a matrix to filter out the redundant values
                \STATE {\bfseries Set} $F \gets \boldsymbol 0$
                \# $F \in \mathbb R^{m \times n}$
                \FOR{$hth := 0$ to $h-1$}
                    \FOR{$wth := 0$ to $w-1$}
                        \IF{$(wth-i) \bmod k = 0$ {\bfseries and} $wth + k \leq w$ {\bfseries and} $(hth-j) \bmod k = 0$ {\bfseries and} $hth + k \leq h$}
                             \STATE $F[hth][wth] \gets 1$
                        \ENDIF
                    \ENDFOR
                \ENDFOR  
                \STATE $\texttt{ct}.T \gets \texttt{cMul}(F, \texttt{ct}.T)$           
                
                \STATE \# To accumulate the intermediate results
                \STATE $\texttt{ct}.I_s \gets \texttt{Add}(\texttt{ct}.I_s, \texttt{ct}.T)$   
            \ENDFOR
        \ENDFOR

        \STATE {\bfseries Return} $\texttt{ct}.I_s$
    \end{algorithmic}
\end{algorithm}

Next, we will show how to make this prototype homomorphic algorithm work efficiently in real-world cases.

\subsection{Encoding Method  for  Convolution Operation}
For simplicity, we assume that the dataset $X \in \mathbb{R}^{m \times f}$ can be encrypted into a single ciphertext $\texttt{ct}.X$, where $m$ is a power of two, all images are grayscale, and each has size $h \times w$. $\texttt{Volley Revolver}$ encodes the dataset as a matrix using the database encoding method \cite{IDASH2018Andrey} and can handle any CNN layer in a single formation. In most cases, $h \times w < f$; in such cases, zero columns are used for padding. Furthermore, $\texttt{Volley Revolver}$ extends the database encoding method \cite{IDASH2018Andrey} with additional operations to represent the dataset matrix $X$ as a three-dimensional structure.

Algorithm \ref{alg:Homomorphic_convolution_operation} provides a feasible and efficient approach to performing secure convolution operations in the HE domain. However, its working assumption (that the size of an image exactly matches the length of the plaintext), a condition that rarely occurs, limits its practicality in real-world scenarios. Moreover, Algorithm \ref{alg:Homomorphic_convolution_operation} can process only one image at a time, since it assumes that a single ciphertext encrypts a single image, resulting in low efficiency for practical applications.

To solve these problems, $\texttt{Volley Revolver}$ performs some simulated operations on the ciphertext $\texttt{ct}.X$ to treat the two-dimensional dataset as a three-dimensional structure. These simulated operations together could simulate the first continual space of the same size as an image of each row of the matrix encrypted in a real ciphertext as a virtual ciphertext that can perform all the HE operations. Moreover, the number of plaintext slots is usually set to a large number and hence a single ciphertext could encrypt several images.  
For example, the ciphertext encrypting the dataset $X \in \mathbb{R}^{m \times f}$ could be used to simulate $m$ virtual ciphertexts $\texttt{vct}_i$ for $1 \le i \le m$, as shown below:
 
\begin{align*}
& Enc
\begin{bmatrix}
 I_{[1][1]}^{(1)} & I_{[1][2]}^{(1)} & \ldots & I_{[h][w]}^{(1)} & 0 & \ldots & 0 \\
 I_{[1][1]}^{(2)} & I_{[1][2]}^{(2)} & \ldots & I_{[h][w]}^{(2)} & 0 & \ldots & 0 \\
 \vdots & \vdots & \ddots & \vdots & \vdots & \ddots & \vdots \\
 I_{[1][1]}^{(m)} & I_{[1][2]}^{(m)} & \ldots & I_{[h][w]}^{(m)} & 0 & \ldots & 0
\end{bmatrix}
\longrightarrow
\\[0.8em]
&
\begin{bmatrix}
 \texttt{vEnc}\!
 \begin{bmatrix}
  I_{[1][1]}^{(1)} & I_{[1][2]}^{(1)} & \ldots & I_{[h][w]}^{(1)}
 \end{bmatrix}\!
  & 0 & \ldots & 0 \\[0.4em]
 \texttt{vEnc}\!
 \begin{bmatrix}
  I_{[1][1]}^{(2)} & I_{[1][2]}^{(2)} & \ldots & I_{[h][w]}^{(2)}
 \end{bmatrix}\!
  & 0 & \ldots & 0 \\
 \vdots & \vdots & \ddots & \vdots \\[0.4em]
 \texttt{vEnc}\!
 \begin{bmatrix}
  I_{[1][1]}^{(m)} & I_{[1][2]}^{(m)} & \ldots & I_{[h][w]}^{(m)}
 \end{bmatrix}\!
  & 0 & \ldots & 0
\end{bmatrix},
\\[1em]
&\text{or}\\[1em]
& Enc
\begin{bmatrix}
 I_{[1][1]}^{(1)} & I_{[1][2]}^{(1)} & \ldots & I_{[h][w]}^{(1)} & 0 & \ldots & 0 \\
 I_{[1][1]}^{(2)} & I_{[1][2]}^{(2)} & \ldots & I_{[h][w]}^{(2)} & 0 & \ldots & 0 \\
 \vdots & \vdots & \ddots & \vdots & \vdots & \ddots & \vdots \\
 I_{[1][1]}^{(m)} & I_{[1][2]}^{(m)} & \ldots & I_{[h][w]}^{(m)} & 0 & \ldots & 0
\end{bmatrix}
\longrightarrow
\\[0.8em]
& Enc
\begin{bmatrix}
 \texttt{vEnc}\!
 \begin{bmatrix}
  I_{[1][1]}^{(1)} & \ldots & I_{[1][w]}^{(1)} \\
  \vdots & \ddots & \vdots \\
  I_{[h][1]}^{(1)} & \ldots & I_{[h][w]}^{(1)}
 \end{bmatrix}\!
  & 0 & \ldots & 0 \\[0.4em]
 \vdots & \vdots & \ddots & \vdots \\[0.4em]
 \texttt{vEnc}\!
 \begin{bmatrix}
  I_{[1][1]}^{(m)} & \ldots & I_{[1][w]}^{(m)} \\
  \vdots & \ddots & \vdots \\
  I_{[h][1]}^{(m)} & \ldots & I_{[h][w]}^{(m)}
 \end{bmatrix}\!
  & 0 & \ldots & 0
\end{bmatrix}.
\end{align*}

Similar to an HE ciphertext, a virtual ciphertext supports virtual HE operations: $\texttt{vEnc}$, $\texttt{vDec}$, $\texttt{vAdd}$, $\texttt{vMul}$, $\texttt{vRescale}$, $\texttt{vBootstrapping}$, and $\texttt{vRot}$. Except for $\texttt{vRot}$, all other virtual operations can be directly inherited from their corresponding HE operations. Specifically, the HE operations $\texttt{Add}$, $\texttt{Mul}$, $\texttt{Rescale}$, and $\texttt{Bootstrapping}$ produce the same corresponding virtual operations: $\texttt{vAdd}$, $\texttt{vMul}$, $\texttt{vRescale}$, and $\texttt{vBootstrapping}$.

The virtual rotation operation $\texttt{vRot}$ differs from the other virtual operations, as it requires two rotation operations on the underlying real ciphertext, as described in Algorithm \ref{alg:vRot}. In practice, we only need to simulate the rotation on virtual ciphertexts to complete the simulation. The virtual rotation operation $\texttt{vRot}(\texttt{ct}, r)$, which rotates all virtual ciphertexts contained in the real ciphertext $\texttt{ct}$ to the left by $r$ positions, yields the following simulation result:

 \begin{align*} 
 &
 \begin{bmatrix}
 	\texttt{vEnc}
 	\begin{mbmatrix}
  	I_{[1][1]}^{(1)}  &   \ldots  &  I_{[r/w][r\%w]}^{(1)}  & 	I_{[(r+1)/w][(r+1)\%w]}^{(1)}  &  \ldots   &  I_{[h][w]}^{(1)}    
 	\end{mbmatrix}        \\
 	\vdots                \\
	\texttt{vEnc}
 	\begin{mbmatrix}
	I_{[1][1]}^{(m)}  &   \ldots  &  I_{[r/w][r\%w]}^{(m)}  & 	I_{[(r+1)/w][(r+1)\%w]}^{(m)}  &  \ldots   &  I_{[h][w]}^{(m)}   
 	\end{mbmatrix}        \\
 \end{bmatrix}  \\
 &   \hspace{5.02cm} \downarrow{\texttt{vRot}(\texttt{ct}, r)}   \\ 
 &
 \begin{bmatrix}
 	\texttt{vEnc}
 	\begin{mbmatrix}
  	I_{[(r+1)/w][(r+1)\%w]}^{(1)}  &  \ldots   &  I_{[h][w]}^{(1)}      &     I_{[1][1]}^{(1)}  &   \ldots  &  I_{[r/w][r\%w]}^{(1)}     
 	\end{mbmatrix}        \\
 	\vdots                \\
	\texttt{vEnc}
 	\begin{mbmatrix}
	I_{[(r+1)/w][(r+1)\%w]}^{(m)}  &  \ldots   &  I_{[h][w]}^{(m)}      &     I_{[1][1]}^{(m)}  &   \ldots  &  I_{[r/w][r\%w]}^{(m)}     
 	\end{mbmatrix}       \\
 \end{bmatrix}.  
\end{align*}


Given two sets of virtual ciphertexts $\texttt{vct}_{[i]}$ and $\texttt{vct}_{[j]}$ that inhabit two ciphertexts $\texttt{ct}_1$ and $\texttt{ct}_2$ respectively, for $1 \le i , j \le m$, the corresponding virtual HE operation $\texttt{vMul}(\texttt{vct}_{[i]}, \texttt{vct}_{[j]})$ results: 
\begin{align*} 
& Enc
\begin{bmatrix}
 \texttt{vEnc}\!\begin{bmatrix} I_{[1][1]}^{(1)} & \ldots & I_{[h][w]}^{(1)} \end{bmatrix} & 0 & \ldots & 0 \\
 \vdots & \vdots & \ddots & \vdots \\
 \texttt{vEnc}\!\begin{bmatrix} I_{[1][1]}^{(m)} & \ldots & I_{[h][w]}^{(m)} \end{bmatrix} & 0 & \ldots & 0
\end{bmatrix}  
\\[0.8em]
&\hspace{2.02cm} \otimes
\\[0.8em]
& Enc
\begin{bmatrix}
 \texttt{vEnc}\!\begin{bmatrix} G_{[1][1]}^{(1)} & \ldots & G_{[h][w]}^{(1)} \end{bmatrix} & 0 & \ldots & 0 \\
 \vdots & \vdots & \ddots & \vdots \\
 \texttt{vEnc}\!\begin{bmatrix} G_{[1][1]}^{(m)} & \ldots & G_{[h][w]}^{(m)} \end{bmatrix} & 0 & \ldots & 0
\end{bmatrix}  
\\[0.8em]
& \hspace{2.02cm} \Downarrow{\texttt{vMul}(\texttt{vct}_{[i]}, \texttt{vct}_{[j]})}
\\[0.8em]
& Enc
\begin{bmatrix}
 \texttt{vEnc}\!\begin{bmatrix} I_{[1][1]}^{(1)} \cdot G_{[1][1]}^{(1)} & \ldots & I_{[h][w]}^{(1)} \cdot G_{[h][w]}^{(1)} \end{bmatrix} & \ldots  \\
 \vdots & \vdots  \\
 \texttt{vEnc}\!\begin{bmatrix} I_{[1][1]}^{(m)} \cdot G_{[1][1]}^{(m)} & \ldots & I_{[h][w]}^{(m)} \cdot G_{[h][w]}^{(m)} \end{bmatrix} & \ldots 
\end{bmatrix}.
\end{align*}

The virtual HE operation $\texttt{vAdd}$ obtains a similar result:

 \begin{align*} 
& \hspace{0.22cm} Enc
\begin{bmatrix}
 \texttt{vEnc}\begin{bmatrix}
  I_{[1][1]}^{(1)}  & \ldots &  I_{[1][w]}^{(1)} \\
  \vdots            & \ddots &  \vdots           \\
  I_{[h][1]}^{(1)}  & \ldots &  I_{[h][w]}^{(1)}  
 \end{bmatrix} & 0 & \ldots & 0 \\
 \vdots & \vdots & \ddots & \vdots \\
 \texttt{vEnc}\begin{bmatrix}
  I_{[1][1]}^{(m)}  & \ldots &  I_{[1][w]}^{(m)} \\
  \vdots            & \ddots &  \vdots           \\
  I_{[h][1]}^{(m)}  & \ldots &  I_{[h][w]}^{(m)}  
 \end{bmatrix} & 0 & \ldots & 0
\end{bmatrix}  
\\[0.8em]
& \hspace{3.02cm} \oplus
\\[0.8em]
& \hspace{0.22cm} Enc
\begin{bmatrix}
 \texttt{vEnc}\begin{bmatrix}
  G_{[1][1]}^{(1)}  & \ldots &  G_{[1][w]}^{(1)} \\
  \vdots            & \ddots &  \vdots           \\
  G_{[h][1]}^{(1)}  & \ldots &  G_{[h][w]}^{(1)}  
 \end{bmatrix} & 0 & \ldots & 0 \\
 \vdots & \vdots & \ddots & \vdots \\
 \texttt{vEnc}\begin{bmatrix}
  G_{[1][1]}^{(m)}  & \ldots &  G_{[1][w]}^{(m)} \\
  \vdots            & \ddots &  \vdots           \\
  G_{[h][1]}^{(m)}  & \ldots &  G_{[h][w]}^{(m)}  
 \end{bmatrix} & 0 & \ldots & 0
\end{bmatrix}  
\\[0.8em]
& \hspace{3.02cm}\Downarrow{\texttt{vAdd}(\texttt{vct}_{[i]}, \texttt{vct}_{[j]})}
\\[0.8em]
& \hspace{0.02cm} Enc
\begin{bmatrix}
 \texttt{vEnc}\begin{bmatrix}
  I_{[1][1]}^{(1)} + G_{[1][1]}^{(1)}  & \ldots &  I_{[1][w]}^{(1)} + G_{[1][w]}^{(1)} \\
  \vdots            & \ddots &  \vdots                                   \\
  I_{[h][1]}^{(1)} + G_{[h][1]}^{(1)}  & \ldots &  I_{[h][w]}^{(1)} + G_{[h][w]}^{(1)}  
 \end{bmatrix} &  \ldots \\
 \vdots & \vdots \\
 \texttt{vEnc}\begin{bmatrix}
  I_{[1][1]}^{(m)} + G_{[1][1]}^{(m)}  & \ldots &  I_{[1][w]}^{(m)} + G_{[1][w]}^{(m)} \\
  \vdots            & \ddots &  \vdots                                   \\
  I_{[h][1]}^{(m)} + G_{[h][1]}^{(m)}  & \ldots &  I_{[h][w]}^{(m)} + G_{[h][w]}^{(m)}  
 \end{bmatrix} &  \ldots 
\end{bmatrix}.
\end{align*}


\begin{algorithm}[htbp]
    \caption{vRot: virtual rotation operations on a set of virtual ciphertexts}
    \label{alg:vRot}
    \begin{algorithmic}[1]
        \STATE {\bfseries Require:} a ciphertext $\texttt{ct}.X$ encrypting a dataset matrix $X$ of size $m \times n$, each row of which encrypts an image of size $h \times w$ such that $h \times w \le n$, and the number $r$ of rotations to the left
        \STATE {\bfseries Ensure:} a ciphertext $\texttt{ct}.R$ encrypting the resulting matrix $R$ of the same size as $X$

        \STATE {\bfseries Set} $R \gets \boldsymbol 0$
        \# $R \in \mathbb R^{m \times n}$
        \STATE $\texttt{ct}.R \gets \texttt{Enc}_\textit{pk}(R)$
        
        \STATE \# Step 1: Rotate the real ciphertext to the left by $r$ positions first and then clean up the garbage values
        \STATE $\texttt{ct}.T \gets \texttt{Rot}(\texttt{ct}.X, r)$
        
        \STATE \# Build a specially designed matrix
        \STATE {\bfseries Set} $F_1 \gets \boldsymbol 0$
        \# $F_1 \in \mathbb R^{m \times n}$
        \FOR{$i := 1$ to $m$}
            \FOR{$j := 1$ to $h \times w - r$}
                \STATE $F_1[i][j] \gets 1$
            \ENDFOR
        \ENDFOR
        \STATE $\texttt{ct}.T_1 \gets \texttt{cMul}(F_1, \texttt{ct}.T)$
          
        \STATE \# Step 2: Rotate the real ciphertext to the right by $h \times w - r$ positions (the same as to the left by $m \times n - h \times w + r$ positions) first and then clean up the garbage values
        \STATE $\texttt{ct}.P \gets \texttt{Rot}(\texttt{ct}.X, m \times n - h \times w + r)$
        
        \STATE \# Build a specially designed matrix 
        \STATE {\bfseries Set} $F_2 \gets \boldsymbol 0$
        \# $F_2 \in \mathbb R^{m \times n}$
        \FOR{$i := 1$ to $m$}
            \FOR{$j := h \times w - r + 1$ to $h \times w$}
                \STATE $F_2[m][j] \gets 1$
            \ENDFOR
        \ENDFOR
        \STATE $\texttt{ct}.T_2 \gets \texttt{cMul}(F_2, \texttt{ct}.P)$
          
        \STATE \# Concatenate            
        \STATE $\texttt{ct}.R \gets \texttt{Add}(\texttt{ct}.T_1, \texttt{ct}.T_2)$
 
        \STATE {\bfseries Return} $\texttt{ct}.R$
    \end{algorithmic}
\end{algorithm}

To bring all the pieces together, we can use  Algorithm \ref{alg:Homomorphic_convolution_operation} to perform convolution operations for several images in parallel based on the simulation virtual ciphertexts. The most efficient part of these simulated operations is that a sequence of operations on a real ciphertext results in the same corresponding operations on the multiple  virtual ciphertexts, which would suffice the real-world applications. 

\paragraph{Limitations} 
Our current framework assumes that a single ciphertext possesses sufficient SIMD capacity (i.e., plaintext slots) to encapsulate at least one complete channel of an (color) image; otherwise, the computational efficiency may be significantly compromised.
In practical scenarios where the pixel count of a high-resolution color channel exceeds the number of available slots in a standard ciphertext---thus violating the single-ciphertext constraint---it becomes necessary to distribute one or more color images across multiple ciphertexts. Our ongoing work demonstrates that our framework is highly versatile and scalable in managing such constraints by employing multiple ciphertexts to represent one or more high-resolution images. Specifically, ongoing follow-up work indicates that our approach can seamlessly scale by utilizing a number of ciphertexts proportional to the image width to encode each channel.


Furthermore, our current framework requires the data owner to expand the convolutional kernels into multiple ciphertexts, each representing a distinct periodic pattern derived from the kernel's spatial expansion. 
Our ongoing follow-up research indicates that this limitation can be effectively mitigated: each kernel needs to be encrypted into only a single ciphertext that encapsulates its full information in an initial expanded form. 
The subsequent transformed representations required for the convolution operation can then be generated on the cloud server through ciphertext rotations. 
This optimization significantly reduces the client-side encryption overhead and the overall communication cost from $O(k^2)$ to $O(1)$ ciphertexts per kernel.



\section{Secure Inference}
In most real-world applications, datasets are typically too large to be encrypted into a single ciphertext. In such cases, the dataset can be partitioned into multiple mini-batches, each containing an equal number of data samples. Each mini-batch is then encrypted into a separate ciphertext. This approach not only allows the continued use of the algorithms described above but also enables improved parallelization.
\subsection{Parallelized Matrix Multiplication}
Algorithm~\ref{alg:Homomorphic_matrix_multiplication}  is designed for matrix multiplication based on the assumption that all data can be encrypted into a single ciphertext. This algorithm can be accelerated through improvements in hardware threading capabilities. Suppose we have two matrices \( A \) of size \( m \times n \) and \( B \) of size \( n \times p \), where the algorithm requires \( O(p \log n) \) ciphertext rotations. If we have \( p \) hardware threads available, then each thread can work in parallel, requiring only \( O(\log p) \) ciphertext rotations per thread.

If the dataset is too large to be encrypted into a single ciphertext, it can be partitioned and encrypted into multiple ciphertexts, represented by two sets, \( A_1, \dots, A_i \) and \( B_1, \dots, B_j \). The proposed algorithm can still be applied under this setup, and the matrix multiplication can be further parallelized at the ciphertext level: each ciphertext in set \( A \) needs to be processed with each ciphertext in set \( B \) and this can be done in parallel. Mathematically, this approach leverages submatrix computations to ensure theoretical consistency and correctness.

This conveys that even with multiple ciphertexts, the algorithm remains applicable and can achieve parallelism by processing each pair of ciphertexts from the two sets, leveraging the computation principles of submatrices.

\subsection{Parallelized Convolution Operation}
With the simulation operations on the real ciphertexts, we can use  Algorithm \ref{alg:Homomorphic_convolution_operation} to perform convolution operations for several images simultaneously on the virtual ciphertexts in parallel. Depending on the number of virtual ciphertexts that can be simulated within a single real ciphertext, the algorithm's time complexity can be further reduced through amortization. For instance, if one real ciphertext can house \( m \) virtual ciphertexts, the algorithm can conduct convolution operations on \( m \) images at the same time. As a result, the time complexity is reduced to \( \frac{1}{m} \) of the normal time complexity for the algorithm.

Recall that the kernel \( K \) should be utilized in advance to generate \( k^2 \) ciphertexts \( K_1, \ldots, K_{k^2} \), each of which encrypts an \( h \times w \) matrix. Assuming there exists a ciphertext \( \texttt{ct}.I \) that encrypts a one-dimensional channel of a gray image of size \( h \times w_I \), the partial convolution computations between the ciphertext \( \texttt{ct}.I \) and the ciphertexts containing kernel information \( K_i \) can be performed in parallel. Moveover, if there are \( k^2 \) hardware-supported threads available, each thread can perform computations in parallel, thereby accelerating the overall convolution operation.

For color images, unlike those from the MNIST dataset, which have three channels, or for images after convolutional layers that have multiple channels, each channel can be encrypted into a separate single ciphertext, and processed with Algorithm~\ref{alg:Homomorphic_convolution_operation}. In most cases, a convolutional layer contains multiple kernels, each of which may have multiple channels, with the number of kernel channels matching the number of channels in the input image. In this scenario, the ciphertexts encrypting the different channels of the same image are multiplied with the ciphertexts encrypting the corresponding channels of a given kernel. The resulting ciphertexts can then be utilized by Algorithm~\ref{alg:Homomorphic_convolution_operation} to compute partial convolution results. A final accumulation of these results via ciphertext addition yields the complete convolution output. The operations for different kernels and the same image can be executed in parallel, enabling the algorithm to achieve high performance through extensive parallelization both within and across ciphertexts.

$\texttt{Volley Revolver}$ can be used to build convolutional neural networks as deep as it needs. However, in this case, the computation time will therefore increase and  bootstrapping  will have to be used to refresh the ciphertext, resulting in more time-consuming.

\subsection{Homomorphic CNN  Evaluation}

An important operation following a convolutional layer is to reshape the data structure within the ciphertext, since the convolution computation performed by $\texttt{SumForConv}$ reduces the image size. As a result, the resulting encrypted image data may be encoded differently from the original $\text{Volley Revolver}$ encoding:  
 \begin{align*} 
 & Enc
 \begin{bmatrix}
  \texttt{vEnc}
 \begin{mbmatrix}
  I_{[1][1]}^{(1)}    &   I_{[1][2]}^{(1)}    &  \ldots &  I_{[3][3]}^{(1)}  
 \end{mbmatrix}     &   0   &  \ldots  & 0   \\
  \vdots    &   \vdots   &  \ddots  & \vdots   \\
 \end{bmatrix} \\ 
 & = Enc
 \begin{bmatrix}
 \texttt{vEnc}
 \begin{mbmatrix}
  I_{[1][1]}^{(1)}    &   I_{[1][2]}^{(1)}    &   I_{[1][3]}^{(1)}  \\
  I_{[2][1]}^{(1)}    &  I_{[2][2]}^{(1)}     &    I_{[2][3]}^{(1)}            \\
  I_{[3][1]}^{(1)}    &  I_{[3][2]}^{(1)}     &  I_{[3][3]}^{(1)}    \\  
 \end{mbmatrix}     &   0   &  \ldots  & 0   \\
  \vdots    &   \vdots   &  \ddots  & \vdots   \\
 \end{bmatrix} 
 \\
 &   \hspace{1.65cm} \big\Downarrow{\texttt{SumForConv}(\cdot,2,2) }  
 \\ 
  &  Enc
 \begin{bmatrix}
 \texttt{vEnc}
 \begin{mbmatrix}
  J_{[1][1]}^{(1)}    &   J_{[1][2]}^{(1)}    &   0  \\
  J_{[2][1]}^{(1)}    &  J_{[2][2]}^{(1)}     &    0            \\
  0    &  0    &  0   \\  
 \end{mbmatrix}     &   0   &  \ldots  & 0   \\
  \vdots    &   \vdots   &  \ddots  & \vdots   \\
 \end{bmatrix} 
 \\
  &   \hspace{1.65cm} \big\Downarrow{\texttt{ReForm}(\cdot,2,2) }  
 \\
  & \ \ \ \ \ Enc
 \begin{bmatrix}
 \texttt{vEnc}
 \begin{mbmatrix}
  J_{[1][1]}^{(1)}    &   J_{[1][2]}^{(1)}      \\
  J_{[2][1]}^{(1)}    &  J_{[2][2]}^{(1)}               \\
 \end{mbmatrix}  & 0   &  \ldots     &   0   &   0   &  \ldots  & 0   \\
  \vdots     & \vdots   &  \ldots     &   \vdots   &   \vdots   &  \ddots  & \vdots \\
 \end{bmatrix}
 \\
  & =Enc
 \begin{bmatrix}
 \texttt{vEnc}
 \begin{mbmatrix}
  J_{[1][1]}^{(1)}    &   J_{[1][2]}^{(1)}   & J_{[2][1]}^{(1)}    &  J_{[2][2]}^{(1)}               \\
 \end{mbmatrix}  & 0   &  \ldots     &   0     \\
  \vdots     & \vdots   &  \ldots     &   \vdots    \\
 \end{bmatrix}
\end{align*} 
For the case of stride $1$, this reshaping operation requires $O(h)$ rotations, $O(h)$ constant multiplications, and $O(h)$ additions, where $h$ denotes the height of the original unprocessed image.

A pooling layer typically would reduce the size of an input image and, similarly, requires reshaping the data structure within the ciphertext. Since the baseline~\cite{jiang2018secure} did not include a pooling layer, we did not apply one either.

Between successive layers (e.g., convolutional layers), the ciphertext's message density is typically reduced and its specific encoding structure is compromised. To ensure inter-layer consistency, the intermediate activations must be reconstructed into a standardized representation to serve as valid input for the subsequent layer. Taking our homomorphic convolution as an example, the raw outputs (as described in the preliminary Algorithm~\ref{alg:Homomorphic_convolution_operation}) require an additional reconstruction phase to restore the original encoding format. For an image matrix of height $H$, this process incurs $O(H)$ rotations to re-align the spatial data. Notably, if the stride $s \ge 2$, the reconstruction complexity increases significantly due to the sparse spatial distribution of the result slots. Consequently, we advocate for the use of unit stride ($s=1$), which is a standard configuration in deep learning architectures and minimizes the rotational overhead required for structural preservation.

\paragraph*{The Threat Model}
If the underlying HE scheme can ensure IND-CPA security, meaning that the ciphertexts of any two messages are computationally indistinguishable, then all computations performed on the public cloud will be over encrypted data. This ensures that the cloud server, which operates in a semi-honest model, learns nothing from the encrypted data. Consequently, we can guarantee the confidentiality of the data against such an adversary.

\paragraph*{The Usage Model}
Our approach can be used in several usage scenarios as illustrated in \cite{han2018efficient}. Whatever the scenario is, the data to be privacy-preserving and the tailored CNN model have to be encrypted before being outsourced to the cloud for its service. In a reasonable usage scenario,  there are a few different roles including data owner,  the model provider, and the cloud server. In some special scenarios, the first two roles can be the same one who would like to get the service from the cloud server. 
Based on this usage model, we can make some plausible assignment preparations 
 for the three roles:
 \begin{enumerate}
 \item Data Owner: The data owner is responsible for preparing the dataset, including tasks such as cleaning and normalizing the data, and, if the dataset is too large, partitioning it into multiple mini-batches of suitable size. Finally, the data owner encrypts the dataset using the database encoding method introduced by \cite{IDASH2018Andrey} and uploads the resulting ciphertext to the cloud.
 
 \item Model Provider: The model provider uses $\texttt{Kernelspanner}$ to generate several ciphertexts encrypting the kernel information, and employs the $\texttt{Volley Revolver}$ method to encode the weight matrix of the fully connected layer and then encrypt the resulting encoded matrix. For the activation layer, the polynomial approximation of the $\texttt{ReLU}$ function can be sent to the cloud as public parameters, without the need for encryption.
 
 \item Cloud Server: The cloud server is responsible for providing cloud-based computation services. To this end, the application implementing the homomorphic CNN algorithm must be deployed on the cloud in advance. In addition, the public keys required for HE operations and the encrypted model weights should be transmitted to the cloud. These preparatory steps can be completed through the collaboration between the model provider and the cloud server.
 \end{enumerate}


\section{Experiments}
Both HE-based and non-HE privacy-preserving frameworks exhibit distinct trade-offs regarding computational overhead and security guarantees. While a comprehensive taxonomic comparison is provided in the baseline study \cite{jiang2018secure}, our analysis focuses strictly on the technical advancements of the proposed method relative to the baseline.



\noindent \textbf{Implementation Details.} To evaluate the efficacy of our framework, we trained custom Convolutional Neural Networks using the TensorFlow ecosystem. The trained model weights were subsequently exported to CSV format for cross-platform compatibility. The core homomorphic inference engine was implemented in C++, leveraging the HEAAN library to perform encrypted operations in the RLWE-based approximate arithmetic setting. To foster reproducibility and support future research, our complete source code is publicly available at the following anonymous repository: \url{https://anonymous.4open.science/r/HE-CNNinfer-ECA4/}.

\subsection{Neural Network Architecture}

We adopt the same CNN architecture as proposed in~\cite{jiang2018secure}, with several modifications to the hyperparameters to optimize the performance of our method. The detailed configuration of our CNN, tailored for the MNIST dataset, is summarized in Table~\ref{tab1}.

\begin{table*}[htbp]
\centering
\caption{Description of our CNN on the MNIST dataset }
\label{tab1}
\begin{tabular}{|l||c|}
\hline
Layer & \multicolumn{1}{c|}{Description}  \\
\hline\hline
CONV               & 32 input images of size $28 \times 28$,  
                            4 kernels of size $3 \times 3$,     
                            stride size of ($1$, $1$)              \\
\hline
ACT-$1$     &  \mysplit{$x \mapsto -0.00015120704 + 0.4610149 \cdot x + 2.0225089 \cdot x^2  -1.4511951 \cdot x^3$}  \\
\hline
FC-$1$      & \mysplit{Fully connected layer with $26 \times 26 \times 4 = 2704$   inputs and $64$ outputs }   \\
\hline
ACT-$2$              &  \mysplit{     $x \mapsto  -1.5650465  -0.9943767 \cdot x  + 1.6794522 \cdot x^2 + 0.5350255 \cdot x^3 $ }    \\
\hline
FC-$2$            &  Fully connected layer with 64 inputs and 10 outputs   \\
\hline
\end{tabular}
\end{table*}

We adopt a standard training pipeline where $\texttt{ReLU}$ is replaced by this polynomial proxy. Specifically, a CNN model (Table~\ref{tab1}) is implemented in $\texttt{Keras}$ and trained on the MNIST dataset (normalized to $[0, 1]$). After achieving $98.66\%$ test accuracy, the pre-trained weights are exported to CSV files. Finally, the model is deployed in the encrypted domain using the $\texttt{HEAAN}$ library, with input pixels scaled by $1/255$ to maintain consistency with the training environment.
Since homomorphic encryption cannot directly evaluate non-polynomial functions like $\texttt{ReLU}$, we approximate it with a degree-three polynomial via the least squares method using $\texttt{Octave}$. This approximation is integrated into all activation layers, with coefficients fine-tuned during training. Core operations, such as matrix multiplications and convolutions, are handled by our proposed HE-efficient algorithms.

\subsection{Homomorphic CNN Inference}
$\textbf{Parameters}.$ Following the notation in \cite{IDASH2018Andrey}, we configure the HE scheme parameters for our implementation as follows: we set the scaling factors to $\Delta = 2^{45}$ for all the inputs and $\Delta_c = 2^{20}$  for the bit precision
of constant values, and use $\texttt{slots} = 32768 = 2^{\texttt{logN}}$. The ciphertext modulus and polynomial modulus degree are configured with $\texttt{logQ} = 1200$ and $\texttt{logN} = 16$, respectively, which collectively provide an estimated $80$-bit security level. See \cite{han2018efficient, jiang2018secure} for further details regarding the parameter selection.

\noindent$\textbf{At the encryption phase}.$  
The data owner encodes the input data $X$ using $\text{Volley Revolver}$ and leverages $\text{SIMD batching}$ to pack multiple images into a single plaintext vector, which is subsequently encrypted into a single ciphertext $ct.I$ under the public key of the Homomorphic Encryption scheme.

On the model-provider side, the process commences with the \texttt{kernelspanner} procedure, which transforms each convolution kernel into a set of discrete ciphertexts, $ct.K_{ij}$ , including the corresponding kernel bias ciphertext $ct.K_b$. For the fully-connected  layers, the weight parameters are transposed into a structured matrix format before being encrypted into weight ciphertexts $ct.W_k$. 

\textit{Encryption of Images.} 
\noindent Under the chosen parameter configuration with $\texttt{slots} = 32{,}768$, the data owner processes $32$ grayscale images of size $28 \times 28$ in a single batch. These images are packed into a plaintext vector and encrypted into a single ciphertext $ct.I$. Consistent with the preprocessing of the training set, MNIST test images are normalized by a factor of $1/255$ prior to encryption. 

\begin{equation*}
X = 
\begin{bmatrix}
 I_{1,1}^{(1)} & \dots & I_{h,w}^{(1)} \\
 \vdots & \ddots & \vdots \\
 I_{1,1}^{(m)} & \dots & I_{h,w}^{(m)} 
\end{bmatrix}
\xrightarrow{\text{Enc}}
\begin{bmatrix}
 I_{1,1}^{(1)} & \dots & I_{h,w}^{(1)} & 0 & \dots & 0 \\
 \vdots & \ddots & \vdots & \vdots & \ddots & \vdots \\
 I_{1,1}^{(m)} & \dots & I_{h,w}^{(m)} & 0 & \dots & 0
\end{bmatrix}.
\end{equation*}

The resulting ciphertext $ct.I$ is subsequently uploaded to and stored on the public cloud in encrypted form.

\textit{Encryption of Trained Model}
For the convolutional layers, each kernel $K_{[i]}$ is expanded via the $\texttt{kernelspanner}$ operation:

\begin{align*}
& K_{[i]} \xrightarrow[\mathbb{R}^{k \times k} \mapsto k^2 \cdot \mathbb{R}^{h \times w}]{\texttt{Kernelspanner}} KI_{[i,j]} 
\xrightarrow[\mathsf{Encrypt}(pk)]{\text{Broadcasting}}  
ct.K_{ij} =
\\
&\quad\quad\quad
\mathsf{Enc} \begin{bmatrix}
 \text{---} & KI_{[i,j]} & \text{---} & 0 & \dots & 0 \\
 \text{---} & KI_{[i,j]} & \text{---} & 0 & \dots & 0 \\
 & \vdots & & \vdots & \ddots & \vdots \\
 \text{---} & KI_{[i,j]} & \text{---} & 0 & \dots & 0
\end{bmatrix},
\\[1em]
& K_{[i,0]} \xrightarrow[\mathbb{R} \mapsto \mathbb{R}^{h \times w}]{\texttt{Kernelspanner}} KB_{[i]} 
\xrightarrow[\mathsf{Encrypt}(pk)]{\text{Broadcasting}} 
ct.K_b =
\\
&\quad\quad\quad
\mathsf{Enc} \begin{bmatrix}
 \text{---} & KB_{[i]} & \text{---} & 0 & \dots & 0 \\
 \text{---} & KB_{[i]} & \text{---} & 0 & \dots & 0 \\
 & \vdots & & \vdots & \ddots & \vdots \\
 \text{---} & KB_{[i]} & \text{---} & 0 & \dots & 0
\end{bmatrix}.
\end{align*}

For the fully connected layers, the weight matrix $W \in \mathbb{R}^{n \times p}$ is transposed and encrypted to ensure compatibility with the ciphertext input:

\begin{align*}
W = 
 \begin{bmatrix}
 w_{1,1}    &  \dots  & w_{1,p}   \\
 \vdots    &  \ddots  & \vdots   \\
 w_{n,1}    &  \dots  & w_{n,p}   \\
 \end{bmatrix}  
\xrightarrow[\mathsf{Encrypt}(pk)]{Encoding(\tau_b)} 
 \mathsf{Enc} \begin{bmatrix}
 w_{1,1}    &   w_{2,1}   &  \dots  & w_{n,1}   \\
 w_{1,2}    &   w_{2,2}   &  \dots  & w_{n,2}   \\
 \vdots    &   \vdots   &  \ddots  & \vdots   \\
 w_{1,p}    &   w_{2,p}   &  \dots  & w_{n,p}   \\
 \end{bmatrix}.
\end{align*}

The structural alignment of the fully connected layers is paramount for maximizing SIMD throughput. For the first FC layer (FC-1), the output dimensionality is $64$, a power of two, allowing our encoding scheme to map the weight matrix directly into SIMD slots without post-transposition padding. Given that the weight matrix ($32 \times 2704$) exceeds the capacity of a single ciphertext, we implement a vertical partitioning strategy. By decomposing the weight matrix along the column dimension into a $32 \times 676 \times 4$ configuration (where $676 = 26 \times 26$), we achieve perfect alignment with the four convolutional output ciphertexts. This structural decomposition yields four corresponding weight sub-matrix ciphertexts, denoted as $ct.W_k$, facilitating efficient homomorphic inner products without necessitating expensive cross-ciphertext permutations. This design ensures that column-wise semantics are preserved, enabling streamlined parallel homomorphic multiplications.

In contrast, the final output layer (FC-2), which performs classification over the $10$ MNIST categories, adopts a power-of-two expansion strategy by extending the output dimension to $16$ nodes through the introduction of $6$ \textit{virtual nodes}. These additional nodes are zero-padded to preserve computational regularity.

The introduction of virtual nodes serves two purposes. First, it simplifies implementation by eliminating the need to explicitly realize the RowShifter algorithm. Second, aligning the dimension to a power of two allows the RowShifter operation—which would otherwise require two ciphertext rotations—to degenerate into a single standard rotation $\mathrm{Rot}(\mathrm{ct}, n)$. As a result, the total number of required rotations is reduced from $2 \times 10$ to only $16$.

For this layer, the $64 \times 16$ weight matrix is transposed and cyclically duplicated to form a $32 \times 64$ matrix, which is subsequently encrypted into a single ciphertext. This structured layout enables optimized SIMD alignment for the homomorphic linear transformation.

\noindent$\textbf{At the prediction phase.}$  
Upon receiving the encrypted image ciphertext $ct.I$ and the encrypted model parameters ($\{ct.K_{ij}\}$, $ct.K_b$, $\{ct.W_k\}$, $ct.V$), the cloud server executes the privacy-preserving CNN inference protocol. The inference process is conducted entirely within the encrypted domain by leveraging homomorphic operations directly on the ciphertexts. By performing these computations without decryption, the server generates encrypted inference results while ensuring the data confidentiality of the owner and the intellectual property of the model provider.

We now describe how each layer of our CNN architecture, as shown in Table~\ref{tab1}, is evaluated under homomorphic encryption.
Since the data owner utilizes SIMD techniques to batch $32$ distinct images, the fully connected layers are formulated as matrix multiplications to facilitate homomorphic evaluation. Specifically, the first FC layer is defined as:
\begin{equation*}
    \mathbb{R}^{32 \times (4\cdot (26 \times 26) )} \times \mathbb{R}^{(4\cdot (26 \times 26) ) \times 64} \longrightarrow \mathbb{R}^{32 \times 64}
\end{equation*}
Similarly, the second FC layer is represented by the transformation:
\begin{equation*}
    \mathbb{R}^{32 \times 64} \times \mathbb{R}^{64 \times 16} \longrightarrow \mathbb{R}^{32 \times 16}
\end{equation*}
This batching strategy allows the cloud server to process all $32$ inference queries simultaneously within the encrypted domain.

\textit{Homomorphic CONV Layer.}  
The cloud server receives the encrypted input image $ct.I$ and the encrypted convolution kernels ($ct.K_{ij}$, $ct.K_b$), where $0 \leq i < 4$ and $0 \leq j < 9$. For each convolution kernel, the server performs homomorphic multiplications between $ct.I$ and $ct.K_{ij}$, followed by the \texttt{SumForConv} operation. This process yields four output ciphertexts $\{ct.C_k\}_{k=0}^3$, with each $ct.C_k$ encrypting the flattened convolution result between the input image $I$ and the $k$-th kernel $K_{[k]}$.

\textit{Activation Layer ACT-1.}  The step applies the degree-3 polynomial function to all the encrypted output images of the homomorphic convolution layer in parallel.
To ensure compatibility with HE, we approximate the ReLU activation function using a degree-3 polynomial, as detailed in Table~\ref{tab1}. This polynomial is applied homomorphically to each ciphertext output of the CONV layer in a component-wise manner.

\textit{Fully connected layer FC-1.}  This procedure perform a matrix multiplication between a $32 \times 2704$ input matrix and a $2704 \times 64$ weight matrix. The output from the activation layer ACT-1 is reshaped into a $32 \times 2704$ input matrix, where each row corresponds to one of the 32 encrypted images. This matrix is multiplied by a $2704 \times 64$ encrypted weight matrix using the above homomorphic matrix multiplication algorithm. To facilitate this, the weight matrix is partitioned into four $32 \times 676$ submatrices, each of which is encrypted into a single ciphertext using the matrix encoding technique introduced in Section~\ref{section:EncodingMethodforMatrixMultiplication}.

\textit{Activation layer ACT-2.}  This step applies the degree-3 polynomial function to all the output nodes of the first FC layer:
The second activation layer again uses a degree-3 polynomial approximation and is applied homomorphically to each of the 64 output nodes from FC-1.

\textit{Fully connected layer FC-2.}  This step performs the multiplication algorithm between  the output ciphertext $ct.S$ of the second activation layer and the weight ciphertext $ct.V$: 
The final inference layer computes the homomorphic product between the encrypted activation output $ct.S$ and the weight  ciphertext $ct.V$. This produces a result ciphertext representing the prediction scores for the $32$ batched images. While the homomorphic evaluation encompasses the full $16$-node augmented output space to preserve SIMD regularity, the inference logic focuses exclusively on the first $10$ slots corresponding to the MNIST categories. The values residing in the $6$ auxiliary slots, generated by the \textit{virtual nodes}, are discarded during the final result extraction. This selective decoding ensures that the classification remains consistent with the original $10$-class problem while benefiting from the performance gains of power-of-two alignment.

\subsection{Performance and Comparison}

We evaluate our implementation of the homomorphic CNN model on the MNIST dataset by calculating ten likelihoods for each of the $32$ encrypted images of handwritten digits, with the homomorphic CNN inference implemented using the \texttt{HEAAN} library \cite{ckks2017homomorphic}.
We evaluate the performance of our implementation on the MNIST testing dataset of 10 000 images. The MNIST database includes a training dataset of 60 000 images and a testing dataset of 10 000, each image of which is of size $28 \times 28$. For such an image, each pixel is represented by a 256-level grayscale and each image depicts a digit from zero to nine and is labeled with it.
 
With the current parameter configuration, our implementation can process $32$ MNIST images simultaneously. Accordingly, the $10{,}000$ MNIST test images are partitioned into $313$ batches, with the final batch zero-padded to form a complete block. Homomorphic CNN inference is then evaluated over these $313$ ciphertext batches, achieving a final classification accuracy of $98.61\%$.
The slight accuracy gap compared to the plaintext model accuracy of $98.66\%$ primarily arises from precision differences. During plaintext training, TensorFlow/Keras employs \texttt{float32} precision, whereas the homomorphic setting operates under reduced numerical precision, with scaling factors $\Delta = 2^{45}$ for inputs and $\Delta_c = 2^{20}$ for constant values.
In practical deployments, this discrepancy can be mitigated through model quantization. For instance, a TensorFlow Lite model can be used to lower plaintext precision from \texttt{float32} to \texttt{float16}, thereby narrowing the precision gap. Alternatively, increasing the scaling factors $\Delta$ and $\Delta_c$ can further improve numerical fidelity in the encrypted domain. Each ciphertext produces 32 classification results (one per image) by outputting the digit with the highest predicted probability, and the processing of a single ciphertext requires $\sim$ 287 seconds on a cloud server equipped with 40 vCPUs. 

The data owner uploads only a single ciphertext ($\sim 19.8$ MB) to the public cloud to encrypt the 32 input images, whereas the model provider must transmit 52 ciphertexts ($\sim 1$ GB) that contain the encrypted weights of the trained model. Upon completion of the  homomorphic evaluation pipeline, the cloud provider transmits a consolidated ciphertext back to a designated trusted authority. This entity, as the sole proprietor of the secret key $\mathsf{sk}$ within the homomorphic encryption scheme, performs the final decryption to recover the computation results. This single-ciphertext transmission minimizes the downlink communication overhead while ensuring that the confidentiality of the processed data remains protected against an untrusted cloud environment.

\subsubsection{Baseline Parameters and Implementation} 
In the baseline configuration, all input data are represented with a bit-precision of $\log p = 30$, while the precision for constant values is maintained at $\log p_c = 30$. To ensure a computational security level of at least $80$ bits, the ring dimension is set to $N = 2^{13}$, resulting in a maximum fresh ciphertext modulus of $\log q \approx 250$. Under these specific parameters, the footprint of a single fresh ciphertext is $0.488$ MB. 

The data owner initiates the process by selecting $64$ images from the MNIST dataset, which are normalized to the range $[0, 1]$ by dividing each pixel value by $255$. These images are subsequently encoded into $7 \times 7 = 49$ ciphertexts. The aggregate communication overhead for the encrypted input is $0.488 \times 49 \approx 23.926$ MB. 

Concurrently, the model provider generates three distinct categories of ciphertexts to encapsulate the model parameters: 
(i) $\{ \mathsf{ct}_{i,j}^{K(k)} \}$ for $0 \leq i,j < 7$ and $0 \leq k < 4$; 
(ii) $\{ \mathsf{ct}_{k}^{W(\ell)} \}$ for $0 \leq k < 4$ and $0 \leq \ell < 64$; and 
(iii) $\{ \mathsf{ct}^{V(\ell)} \}$ for $0 \leq \ell < 16$. 
The total storage requirement for these $468$ model ciphertexts is $0.488 \times 468 \approx 228.516$ MB.

The end-to-end homomorphic classification of the encrypted batch requires approximately $1.69$ seconds. This baseline model achieves a classification accuracy of $98.1\%$ on the test set, matching the performance observed in the plaintext domain (i.e., evaluation in the clear). This equivalence demonstrates that the error-introduction inherent in approximate homomorphic encryption does not result in a non-negligible loss of inference precision.

\subsubsection{Comparison with the Baseline} Compared to the baseline configuration, the primary advantages of our proposed framework are summarized as follows:

\begin{itemize}
    \item \textbf{Optimal Ciphertext Packing:} 
    Our scheme achieves very high slot utilization. By using the minimum required number of ciphertexts, we maximize SIMD packing efficiency, ensuring that every plaintext slot is effectively used for computation. In practical scenarios where the number of image pixels is a power of two, our method can achieve a \textit{zero-slot-waste} configuration.  
    
    Under the parameter setting adopted in this paper, each MNIST image must be packed into a slot capacity of $1024$, which exceeds its native resolution of $28 \times 28$. As a result, the slot utilization ratio is $28 \times 28 / 1024 = 76.56\%$. In practical deployments, MNIST images can be resized to $32 \times 32$ to fully occupy all available slots, achieving $100\%$ utilization and potentially improving model performance. 
    
    Our encoding method also naturally extends to encrypted multi-channel image data. Consider, for example, a batch of $32$ color images of size $32 \times 32$ with three channels. Under the same parameter configuration as used in this paper, our method requires only three ciphertexts to encrypt the entire batch—one ciphertext per channel. Each ciphertext packs the corresponding channel values across all $32$ images in the same manner as in the MNIST grayscale setting. In this case, all slots in the three ciphertexts are fully utilized, resulting in an optimal packing configuration with zero slot waste.
    
    \item \textbf{Scalable Hardware Acceleration: }
    Although ciphertext rotations incur non-trivial computational latency, they are level-neutral and do not deplete the multiplicative depth of the homomorphic scheme. By prioritizing rotations over depth-intensive operations, our framework enhances parallel scalability without accelerating the consumption of the modulus chain. This design strategically defers the necessity of costly bootstrapping (ciphertext refreshing) operations, which are significantly more resource-intensive and time-consuming than rotations. Consequently, the quadratic gradient framework ensures a more sustainable level-budget management for complex optimization trajectories. 

    Unlike multiplicative operations that rapidly exhaust the circuit depth, ciphertext rotations facilitate high-performance SIMD parallelism without sacrificing the remaining levels of the ciphertext. By trading off increased rotation counts for a preserved multiplicative budget, we minimize the frequency of performance-bottlenecking bootstrapping stages. This approach is particularly advantageous for large-scale HE workloads, as rotations can be efficiently offloaded to parallel hardware accelerators (e.g., GPUs or FPGAs), effectively neutralizing the individual operation latency while reaping the benefits of massive parallel scalability. 

    Our framework is architected such that the primary matrix computations and convolutional primitives are executed in a highly parallelized manner. By leveraging the SIMD properties of the HE scheme, we encode multiple image instances into a single ciphertext, ensuring that each ciphertext encapsulates semantically-complete information for the respective data samples. This design significantly facilitates the construction of parallelized evaluation algorithms. We observe that the system's inference latency is heavily contingent upon the available computational resources, specifically the number of vCPUs provisioned in the cloud environment. To characterize this sensitivity, we initially evaluated our artifact on a server equipped with $12$ vCPUs, where the classification of $32$ MNIST images yielded a response time of approximately $30$ minutes. In contrast, our primary experimental results were obtained using a high-performance configuration with $40$ vCPUs. Under this setup, the end-to-end execution time was reduced to $285$ seconds. This performance gain underscores the scalability of our approach and highlights the trade-off between hardware resource allocation and computational throughput in privacy-preserving deep learning.
    
    \item \textbf{Multidimensional Tensor Architecture: } 
    Unlike prior arts that necessitate flattening multidimensional structures into 1D or 2D mappings, our framework introduces a 3D tensor-native architecture building upon state-of-the-art database encoding techniques \cite{IDASH2018Andrey}. This design preserves the intrinsic structural relationships of Deep Learning tensors, making it highly compatible with both privacy-preserving inference and training. Ongoing research further demonstrates the efficacy of this encoding in HE-based neural network training. 

    Our proposed encoding scheme is specifically tailored for deep learning architectures characterized by stacked convolutional layers with multiple filters operating on multi-channel color imagery. While subsequent extensions of prior baseline works have demonstrated scalability to multi-layer benchmarks utilizing only a \textit{single filter per layer}, their generalizability to more complex, multi-filter architectures remains an open research question. 

    To bridge this gap, we address the non-trivial challenge of representing the standard four-dimensional (4D) input tensor $\mathcal{T} \in \mathbb{R}^{B \times C \times H \times W}$ within the algebraic constraints of cryptographic primitives. Here, the dimensions represent the batch size ($B$), the number of input channels ($C$), and the spatial resolution (height $H$ and width $W$), respectively.
    Although our construction encodes a 3D tensor structure (i.e., $B \times H \times W$) within a single ciphertext, we effectively handle the fourth dimension (e.g., the channel dimension) by mapping it across distinct ciphertexts. This approach faithfully preserves the underlying data flow semantics of Convolutional Neural Networks. Consequently, our framework provides native support for both raw multi-channel datasets (e.g., CIFAR-10) and the intermediate feature maps (i.e., convolved activations) generated throughout the inference pipeline.
\end{itemize}

\noindent \textbf{Computational Overheads and Latency:} 
Despite the optimization in packing efficiency, our framework exhibits higher inference latency compared to the baseline. This is primarily attributed to the fact that our scheme requires a higher frequency of ciphertext rotations, which are significantly more computationally expensive than homomorphic additions or multiplications.

Moreover, there exists a notable trade-off between the baseline and our proposed framework in terms of computational depth. Specifically, our algorithm inherently incurs a greater multiplicative depth, thereby requiring larger encryption parameters (i.e., higher $\log N$ and $\log Q$) to accommodate cumulative noise growth throughout the inference pipeline. This increased depth, together with more complex tensor-native operations, contributes substantially to the higher execution latency relative to the baseline.

Furthermore, our design adopts a larger polynomial modulus degree ($N$) and a correspondingly larger ciphertext modulus ($Q$), which are standard configurations for maintaining strong security guarantees while supporting deeper homomorphic computations. Although a larger $\log Q$ enables more complex evaluations, it inevitably increases the execution cost of primitive homomorphic operations.

In addition, the baseline employs a quadratic activation (degree-2), whereas our model uses a degree-3 polynomial approximation for all activation layers (ACT-1 and ACT-2). While the cubic polynomial provides a more accurate approximation to the ReLU function, it also results in a deeper homomorphic circuit and thus higher computational overhead.

Finally, although the proposed architecture is inherently parallelizable across multiple convolution kernels, our current implementation does not yet fully exploit this concurrency (currently limited to 40 vCPUs). In practice, the four convolution kernels are executed sequentially, accounting for a substantial portion of the total response time (approximately 196 seconds). We anticipate that executing these convolution kernels in parallel when deployed on high-performance servers would significantly mitigate these latency bottlenecks.

\noindent \textit{Throughput Scalability.} It is worth noting that under the baseline parameter setting of $\log N = 13$, our proposed framework demonstrates a significant advantage in batching efficiency. Specifically, our scheme enables each individual ciphertext to independently encrypt 8 complete images. In contrast, the baseline framework requires a cooperative encryption approach, where 49 ciphertexts must work in tandem to represent a batch of 64 images. Consequently, for a fixed allocation of 49 ciphertexts, our framework can accommodate up to 392 images ($8 \times 49$), achieving a $6.125\times$ throughput improvement over the state-of-the-art.

However, this gain comes with a trade-off in multiplicative depth. Our algorithm consumes more modulus budget than the baseline, and under this configuration, a ciphertext modulus setting of $\log Q = 250$ may be insufficient to support the full inference circuit without additional noise management. In such cases, ciphertext refreshing (bootstrapping) would be required to sustain correct decryption.

\section{Conclusion}
The encoding method we proposed in this work,   $\texttt{Volley Revolver}$, is particularly tailored for privacy-preserving neural networks. There is a great chance that it can be used to assist the private neural networks training, in which case for the backpropagation algorithm of the fully-connected layer the first matrix $A$ is revolved while the second matrix $B$ is settled to be still. Ongoing research clearly shows that our scheme is highly suitable for FHE-based neural network training.

We shifted some work related to the CNN model to the model provider and some data preparation processes to the data owner so as to complete the homomorphic CNN inference. We believe it is all right for privacy-preserving inference due to no sensitive information leaking.

In this work, we assume that the images are grayscale and that the image dataset can be encrypted into a single ciphertext. Our method, however, is also applicable to large-scale color image datasets and can accommodate different stride sizes while using the same padding technique. Ongoing research indicates that it can be extended to real-world applications involving high-resolution color images.

\section{Acknowledgments}

The authors utilized AI-based assistance, including Grammarly \cite{grammarly}, OpenAI ChatGPT \cite{chatgpt}, and Google Gemini \cite{gemini}, to rectify typographical errors, improve grammatical structure, and refine awkward phrasing throughout the manuscript.

\bibliographystyle{ACM-Reference-Format}
\bibliography{HE.CNNinfer}

@article{grammarly,
  title={},
  author={Grammarly Inc.},
  journal={},
  url = "https://www.grammarly.com/",
  pages={},
  year={2024}
}

@article{chatgpt,
  title={ChatGPT},
  author={OpenAI},
  journal={},
  url = "https://openai.com/chatgpt/",
  pages={},
  year={2024}
}

@misc{gemini,
  author = {Google},
  title = {Gemini},
  year = {2024},
  url = {https://gemini.google.com/}
}

@inproceedings{barni2006privacy,
  title={A privacy-preserving protocol for neural-network-based computation},
  author={Barni, Mauro and Orlandi, Claudio and Piva, Alessandro},
  booktitle={Proceedings of the 8th workshop on Multimedia and security},
  pages={146--151},
  year={2006}
}

@article{orlandi2007oblivious,
  title={Oblivious neural network computing via homomorphic encryption},
  author={Orlandi, Claudio and Piva, Alessandro and Barni, Mauro},
  journal={EURASIP Journal on Information Security},
  volume={2007},
  number={1},
  pages={037343},
  year={2007},
  publisher={Springer}
}

@inproceedings{demmler2015aby,
  title={ABY-A framework for efficient mixed-protocol secure two-party computation.},
  author={Demmler, Daniel and Schneider, Thomas and Zohner, Michael},
  booktitle={Ndss},
  year={2015}
}

@inproceedings{liu2017oblivious,
  title={Oblivious neural network predictions via minionn transformations},
  author={Liu, Jian and Juuti, Mika and Lu, Yao and Asokan, Nadarajah},
  booktitle={Proceedings of the 2017 ACM SIGSAC conference on computer and communications security},
  pages={619--631},
  year={2017}
}

@inproceedings{riazi2018chameleon,
  title={Chameleon: A hybrid secure computation framework for machine learning applications},
  author={Riazi, M Sadegh and Weinert, Christian and Tkachenko, Oleksandr and Songhori, Ebrahim M and Schneider, Thomas and Koushanfar, Farinaz},
  booktitle={Proceedings of the 2018 on Asia conference on computer and communications security},
  pages={707--721},
  year={2018}
}

@inproceedings{juvekar2018gazelle,
  title={$\{$GAZELLE$\}$: A low latency framework for secure neural network inference},
  author={Juvekar, Chiraag and Vaikuntanathan, Vinod and Chandrakasan, Anantha},
  booktitle={27th USENIX security symposium (USENIX security 18)},
  pages={1651--1669},
  year={2018}
}

@inproceedings{bourse2018fast,
  title={Fast homomorphic evaluation of deep discretized neural networks},
  author={Bourse, Florian and Minelli, Michele and Minihold, Matthias and Paillier, Pascal},
  booktitle={Annual International Cryptology Conference},
  pages={483--512},
  year={2018},
  organization={Springer}
}

@inproceedings{gilad2016cryptonets,
  title={Cryptonets: Applying neural networks to encrypted data with high throughput and accuracy},
  author={Gilad-Bachrach, Ran and Dowlin, Nathan and Laine, Kim and Lauter, Kristin and Naehrig, Michael and Wernsing, John},
  booktitle={International conference on machine learning},
  pages={201--210},
  year={2016},
  organization={PMLR}
}

@inproceedings{mohassel2017secureml,
  title={Secureml: A system for scalable privacy-preserving machine learning},
  author={Mohassel, Payman and Zhang, Yupeng},
  booktitle={2017 IEEE symposium on security and privacy (SP)},
  pages={19--38},
  year={2017},
  organization={IEEE}
}

@misc{SmartandVercauteren_SIMD,
    author       = {N.P. Smart and
		    F. Vercauteren},
    title        = {Fully Homomorphic SIMD Operations},
    howpublished = {Cryptology ePrint Archive, Report 2011/133},
    year         = {2011},
    note         = {\url{https://ia.cr/2011/133}},
}

@article{halevi2020helib,
  title={HElib design principles},
  author={Halevi, Shai and Shoup, Victor},
  journal={Tech. Rep.},
  year={2020},
  note = {\url{https://github.com/homenc/HElib}}
}

@article{chou2018faster,
  title={Faster cryptonets: Leveraging sparsity for real-world encrypted inference},
  author={Chou, Edward and Beal, Josh and Levy, Daniel and Yeung, Serena and Haque, Albert and Fei-Fei, Li},
  journal={arXiv preprint arXiv:1811.09953},
  year={2018}
}

@article{chabanne2017privacy,
  title={Privacy-Preserving Classification on Deep Neural Network.},
  author={Chabanne, Herv{\'e} and de Wargny, Amaury and Milgram, Jonathan and Morel, Constance and Prouff, Emmanuel},
  journal={IACR Cryptol. ePrint Arch.},
  volume={2017},
  pages={35},
  year={2017}
}

@inproceedings{brutzkus2019low,
  title={Low latency privacy preserving inference},
  author={Brutzkus, Alon and Gilad-Bachrach, Ran and Elisha, Oren},
  booktitle={International Conference on Machine Learning},
  pages={812--821},
  year={2019},
  organization={PMLR}
}

@inproceedings{jiang2018secure,
  title={Secure outsourced matrix computation and application to neural networks},
  author={Jiang, Xiaoqian and Kim, Miran and Lauter, Kristin and Song, Yongsoo},
  booktitle={Proceedings of the 2018 ACM SIGSAC Conference on Computer and Communications Security},
  pages={1209--1222},
  year={2018}
}

@inproceedings{gentry2009fully,
  title={Fully homomorphic encryption using ideal lattices},
  author={Gentry, Craig},
  booktitle={Proceedings of the forty-first annual ACM symposium on Theory of computing},
  pages={169--178},
  year={2009}
}

@article{IDASH2018Andrey,
  title={Logistic regression model training based on the approximate homomorphic encryption},
  author={Kim, Andrey and Song, Yongsoo and Kim, Miran and Lee, Keewoo and Cheon, Jung Hee},
  journal={BMC medical genomics},
  volume={11},
  number={4},
  pages={83},
  year={2018},
  publisher={Springer}
}

@inproceedings{han2018efficient,
  title={Logistic regression on homomorphic encrypted data at scale},
  author={Han, Kyoohyung and Hong, Seungwan and Cheon, Jung Hee and Park, Daejun},
  booktitle={Proceedings of the AAAI Conference on Artificial Intelligence},
  volume={33},
  number={01},
  pages={9466--9471},
  year={2019},
  
  doi = "https://doi.org/10.1609/aaai.v33i01.33019466"
}

@inproceedings{ckks2017homomorphic,
  title={Homomorphic encryption for arithmetic of approximate numbers},
  author={Cheon, Jung Hee and Kim, Andrey and Kim, Miran and Song, Yongsoo},
  booktitle={International Conference on the Theory and Application of Cryptology and Information Security},
  pages={409--437},
  year={2017},
  organization={Springer}
}

@String{Computing = "Computing" }

@String{Computer = "{IEEE} Computer" }

@String{Springer = "Springer-Verlag" }

@ArtifactSoftware{R,
    title = {R: A Language and Environment for Statistical Computing},
    author = {{R Core Team}},
    organization = {R Foundation for Statistical Computing},
    address = {Vienna, Austria},
    year = {2019},
    url = {https://www.R-project.org/},
}

\end{document}